\documentclass[12pt,epsf,amssymb]{article} \textheight =23 truecm
\def\lsim{\raise0.3ex\hbox{$<$\kern-0.75em\raise-1.1ex\hbox{$\sim$}}}
\def\gsim{\raise0.3ex\hbox{$>$\kern-0.75em\raise-1.1ex\hbox{$\sim$}}}
\def\noi{\noindent} \def\nn{\nonumber} \def\bea{\begin{eqnarray}}
\def\eea{\end{eqnarray}} \def\beq{\begin{equation}}
\def\eeq{\end{equation}} 
\def\beeq{\begin{eqnarray}} \def\eeeq{\end{eqnarray}} \def\R{ {\rm R
\kern -.31cm I \kern .15cm}} \def\C{ {\rm C \kern -.15cm \vrule
width.5pt \kern .12cm}} \def\Z{ {\rm Z \kern -.27cm \angle \kern
.02cm}} \def\N{ {\rm N \kern -.26cm \vrule width.4pt \kern .10cm}}
\def\1{{\rm 1\mskip-4.5mu l} }

\usepackage{graphics} \usepackage{graphicx} \usepackage{epsfig}

\begin{document} \begin{center} {\Large \bf Finite mass corrections for $B \to (\overline{D}^{(*)}, \overline{D}^{**}) \ell \nu$ decays in the Bakamjian-Thomas relativistic quark model} \\

\vskip 1 truecm {\bf H.-R. Dong}\\

{\it Institute of High Energy Physics IHEP, Chinese Academy of Sciences}\\
{\it Theoretical Physics Center for Science Facilities TPCSF}\\
{\it 196 Yuquan Lu, Shijingshan district, 100049 Beijing, China} 

\vskip 0.5 truecm {\bf A. Le Yaouanc, L. Oliver and J.-C. Raynal}\\
{\it Laboratoire de Physique Th\'eorique}\footnote{Unit\'e Mixte de
Recherche UMR 8627 - CNRS }\\    {\it Universit\'e de Paris XI,
B\^atiment 210, 91405 Orsay Cedex, France} \end{center}

\vskip 0.5 truecm

\begin{abstract}

The Bakamjian-Thomas relativistic quark model for hadron current matrix elements, while non-covariant at finite mass, is successful in the heavy quark limit : form factors are covariant and satisfy Isgur-Wise scaling and Bjorken-Uraltsev sum rules. Motivated by the so-called "$1/2$ vs. $3/2$ puzzle" in $\overline{B}$ decays to positive parity $D^{**}$, we examine the implications of the model at finite mass. In the elastic case ${1 \over 2}^- \to {1 \over 2}^-$, the HQET constraints for the $O(1/m_Q)$ corrections are analytically fulfilled. A number of satisfying regularities is also found for inelastic transitions. We compute the form factors using the wave functions given by the Godfrey-Isgur potential.  We find a strong enhancement in the case ${1 \over 2}^- \to {1 \over 2}^+$ for $0^- \to 0^+$. This enhancement is linked to a serious difficulty of the model at finite mass for the inelastic transitions, namely a violation of the HQET constraints at zero recoil formulated by Leibovich et al. These are nevertheless satisfied in the non-relativistic limit for the light quark. We conclude that these HQET rigorous constraints are crucial in the construction of a sensible relativistic quark model of inelastic form factors.\par

\end{abstract}

\vskip 0.2 truecm

\noi LPT-Orsay-14-40 \qquad \qquad July 2014

\par \vskip 0.2 truecm

\noindent e-mails :  donghr@ihep.ac.cn, Alain.Le-Yaouanc@th.u-psud.fr, Luis.Oliver@th.u-psud.fr 

\newpage \pagestyle{plain}

\section{Introduction} \hspace*{\parindent}

The Bakamjian-Thomas (BT) relativistic quark models \cite{BT,KP,T,CGNPSS} are a class of models with a fixed number of constituents in which the states are covariant under the Poincar\'e group. The model relies on an appropriate Lorentz boost of the eigenfunctions of a Hamiltonian describing the hadron spectrum at rest.\par 

We have proposed a formulation of this scheme for the meson ground states \cite{LOPR-1} and demonstrated the important feature that, in the heavy quark limit, the current matrix elements, when the current is coupled to the heavy quark, are {\it covariant}. We have extended this scheme to P-wave excited states \cite{MLOPR-1}.\par

Moreover, these matrix elements in the heavy quark limit exhibit Isgur-Wise (IW) scaling \cite{IW-1}. As demonstrated in \cite{LOPR-1, MLOPR-1}, given a Hamiltonian describing the spectrum, the model provides an unambiguous result for the Isgur-Wise functions, the elastic $\xi (w)$ \cite{IW-1} and the inelastic to P-wave states $\tau_{1/2}(w)$, $\tau_{3/2}(w)$ \cite{IW-2}.\par 

On the other hand, the sum rules (SR) in the heavy quark limit of QCD, like Bjorken \cite{BJ,IW-2} and Uraltsev SR \cite{UR} are analytically satisfied in the model \cite{LOPR-2,MLOPR-2,LOPRM}, as well as SR involving higher derivatives of $\xi (w)$ at zero recoil \cite{LOR-1,LOR-2,LOR-3}.\par
  
In \cite{MLOPR-3}, we have chosen the Godfrey-Isgur Hamitonian \cite{GI}, that gives a very complete description of the light $q\overline{q}$ and heavy $Q\overline{q}$ meson spectra in order to predict within the BT scheme the corresponding IW functions for the ground state and the excited states.\par

Similar work has been been performed for $Q\overline{q}$ meson decay constants \cite{MLOPR-4} and to demonstrate within the BT scheme new Heavy Quark Effective Theory (HQET) SR involving Isgur-Wise functions and decay constants \cite{LOPR-3}.\par  

A detailed and very useful account of the BT scheme for the calculation of Isgur-Wise functions and heavy meson decay constants and their numerical calculation within the Godfrey-Isgur Hamiltonian has been given in the PhD Thesis of Vincent Mor\'enas \cite{M}.

As a further test, we have computed in \cite{BCLOR}, the vector, scalar and axial charge densities for the ground states $0^-$ and $1^-$ (${1 \over 2}^-$ doublet) and for the excited states $0^+$ and $1^+$ (${1 \over 2}^+$ doublet). In this case the active quark is the light quark, and one can show that, unlike the case of the active heavy quark, the current matrix elements are not covariant. For the calculation, we have adopted the natural reference frame for this problem, the heavy meson rest frame. As shown in \cite{BCLOR}, the agreement with lattice data in the unquenched approximation is really striking, and provides both a test of the BT scheme and of the GI Hamiltonian that describes the spectrum.\par

A main motivation to undertake this work has been the so-called "${1 \over 2}$ versus ${3 \over 2}$ puzzle" that, based on rather old data, states the fact that the semileptonic decay rates ${1 \over 2}^- \to {1 \over 2}^+$ are much larger than the expectations of the heavy quark limit, while the semileptonic decay rates ${1 \over 2}^- \to {3 \over 2}^+$ are roughly consistent with this limit. A precise discussion of this puzzle has been done in ref. \cite{BIGI}. Updated data by BaBar \cite{BABAR D** SL} and Belle \cite{BELLE D** SL} confirm the problem, although there are significant differences between both experiments.\par 
The ${1 \over 2}$ vs. ${3 \over 2}$ puzzle is nicely exemplified by the Uraltsev Sum Rule \cite{UR} :
\bea
\label{1.1e}
\sum_n \left(|\tau_{3/2}^{(n)} (1)|^2 - |\tau_{1/2}^{(n)} (1)|^2 \right) = {1 \over 4}
\eea

If one neglects completely higher excitations and the ground state ($n = 0$) dominates the sum of the {\it differences} of the l.h.s. of (\ref{1.1e}), one expects $|\tau_{3/2}^{(0)} (1)|^2 > |\tau_{1/2}^{(0)} (1)|^2$. In addition, the phase space factors make much larger the BR for ${1 \over 2}^- \to {3 \over 2}^+$ relatively to the ${1 \over 2}^- \to {1 \over 2}^+$ one. The BT model satisfies analytically \cite{LOPRM} the SR (\ref{1.1e}) with, for $n = 0$ \cite{MLOPR-3} :
\bea
\label{1.2e}
\tau_{1/2}(1) = 0.22 \qquad \qquad \qquad \qquad \tau_{3/2}(1) = 0.54
\eea

On the other hand, calculations in the lattice in the unquenched approximation \cite{BLOSSIER} point to a similar conclusion 
\bea
\label{1.2bise}
\tau_{1/2}(1) = 0.29 \pm 0.03 \qquad \qquad \qquad \qquad \tau_{1/2}(1) = 0.52 \pm 0.03
\eea

Let us finally underline that the ${1 \over 2}$ vs. ${3 \over 2}$ puzzle does not seem to be present, assuming factorization, in the nonleptonic decays $\overline{B} \to D^{**} \pi$, as shown by the Belle results \cite{BELLE D**PI}, phenomenologically analyzed in ref. \cite{JLOR}. This feature makes the puzzle even more obscure. Recently, in ref. \cite{LYP} has been done a necessary, precise and updated discussion of the situation for both the semileptonic and nonleptonic data.

The paper is organized as follows. In Section 2 we give the definitions of the form factors for the transitions on which we are interested, reproducing some needed results at leading and $O(1/m_Q)$ order within HQET. In Section 3 we give the master formulae defining the theoretical framework of BT quark models. Since the current matrix elements in the BT model are only covariant in the heavy quark limit if the current is coupled to the heavy quark, the calculation of the $1/m_Q$ corrections must be done in a particular reference frame. We discuss this problem in Section 4 and give arguments to adopt the Equal Velocity Frame (EVF), where the moduli of the initial and final three-vector meson velocities are equal. In Section 5 we check that this frame allows to obtain very reasonable results for the $1/m_Q$ corrections for the elastic transitions ${1 \over 2}^- \to {1 \over 2}^-$. In Section 6 we give the analytical results of the BT model for the $O(1/m_Q)$ of form factors to excited states $\overline{B} \to D^{**} \ell \nu$ at zero recoil, and compare to the results of HQET. Section 7 is devoted to the description of the Godfrey-Isgur quark model for spectroscopy. In Section 8 we give the results of the BT model for the ${1 \over 2}^- \to {1 \over 2}^-$ in the heavy quark limit, at finite mass and at the order $1/m_Q$. Section 9 is devoted to the calculation of the different form factors for the inelastic transitions ${1 \over 2}^- \to {1 \over 2}^+$ and ${1 \over 2}^- \to {3 \over 2}^+$ at infinite and finite mass. In Section 10 we give the numerical results for the branching ratios $\overline{B} \to D^{(*)} \ell \nu, D^{(*)}\pi$ and $\overline{B} \to D^{**} \ell \nu, D^{**}\pi$ in the heavy mass limit and also at finite mass, and in Section 11 we expose a discussion of the obtained results and problems. We leave a number of technicalities to the Appendices. In Appendix A we write the needed formulas of the different form factors in terms of matrix elements. In Appendices B and C we give the wave functions in the GI model, respectively in the heavy quark limit and at finite mass. In Appendix D we write some formulas defining a family of collinear frames and in Appendix E we give the formulas for the decay rates in the different cases.

\section{Matrix elements for $B \to D^{(*)} \ell \nu$ and $B \to \overline{D}^{**} \ell \nu$}

For the ground state mesons $D(0^-)$ and $D^*(1^-)$ we adopt the notation of \cite{FN} :
\bea
\label{3.14e}
{<D(v')|V^\mu|B(v)> \over \sqrt{m_Bm_D}}\ = h_+(w)(v+v')^\mu + h_-(w)(v-v')^\mu
\eea
\bea
\label{3.15e}
{<D^*(v',\epsilon')|V^\mu|B(v)> \over \sqrt{m_Bm_{D^*}}}\ = ih_V(w) \epsilon^\mu_{\ \nu\alpha\beta}\epsilon'^{*\nu} v'^\alpha v^\beta
\eea
\bea
\label{3.16e}
{<D^*(v',\epsilon')|A^\mu|B(v)> \over \sqrt{m_Bm_{D^*}}}\ = h_{A_1}(w)(w+1)\epsilon'^{*\mu}-h_{A_2}(w)(\epsilon'^*.v)v^\mu-h_{A_3}(w) (\epsilon'^*.v)v'^\mu
\eea

\noindent while for the excited P-wave mesons, $D_{1/2}(0^+)$, $D_{1/2}(1^+)$, $D_{3/2}(1^+)$ and $D_{3/2}(2^+)$, we adopt the notation of \cite{LLSW} for the form factors :
\bea
\label{3.17e}
{<D_{3/2}(1^+)(v',\epsilon')|A^\mu|B(v)> \over \sqrt{m_Bm_{D^{**}}}}\ = if_A(w) \epsilon^\mu_{\ \alpha\beta\gamma}\epsilon'^{*\alpha} v'^\beta v'^\gamma
\eea
\bea
\label{3.18e}
{<D_{3/2}(1^+)(v',\epsilon')|V^\mu|B(v)> \over \sqrt{m_Bm_{D^{**}}}}\ = f_{V_1}(w)\epsilon'^{*\mu}+(\epsilon'^*.v)[f_{V_2}(w) v^\mu + f_{V_3}(w) v'^\mu]
\eea
\bea
\label{3.19e}
{<D_{3/2}(2^+)(v',\epsilon')|V^\mu|B(v)> \over \sqrt{m_Bm_{D^{**}}}}\ = ik_V(w) \epsilon'^\mu_{\ \alpha \beta \gamma} \epsilon'^{*\alpha}_\sigma v^\sigma v^\beta v'^\gamma
\eea
\bea
\label{3.20e}
{<D_{3/2}(2^+)(v',\epsilon')|A^\mu|B(v)> \over \sqrt{m_Bm_{D^{**}}}}\ = k_{A_1}(w) \epsilon'^{*\mu}_\alpha v^\alpha + \epsilon'^*_{\alpha\beta} v^\alpha v^\beta [k_{A_2}(w) v^\mu + k_{A_3}(w) v'^\mu]
\eea
\bea
\label{3.21e}
{<D_{1/2}(0^+)(v')|A^\mu|B(v)> \over \sqrt{m_Bm_{D^{**}}}}\ = g_+(w)(v+v')^\mu + g_-(w)(v-v')^\mu
\eea
\bea
\label{3.22e}
{<D_{1/2}(1^+)(v',\epsilon')|A^\mu|B(v)> \over \sqrt{m_Bm_{D^{**}}}}\ = ig_A(w) \epsilon^\mu_{\ \alpha\beta\gamma}\epsilon^{*\alpha} v^\beta v'^\gamma
\eea
\bea
\label{3.23e}
{<D_{1/2}(1^+)(v',\epsilon')|V^\mu|B(v)> \over \sqrt{m_Bm_{D^{**}}}}\ = g_{V_1}(w)\epsilon'^{*\mu}+(\epsilon'^*.v)[g_{V_2}(w)v^\mu+g_{V_3}(w)v'^\mu]
\eea

\noindent In the equations for the excited states $D^{**}$ denotes generically any excited state, but in each equation the physical mass of the corresponding excited meson is understood.

\subsection{Heavy quark expansion of form factors in HQET}

\subsubsection{Elastic form factors $B \to D^{(*)} \ell \nu$ in HQET}

To compare with the results of the BT model at finite mass, let us give here the expressions of the form factors in powers of ${1 \over m_Q}$ in HQET. Let us set the notation $\epsilon_Q = {1 \over {2 m_Q}}$. To first order in the heavy quark expansion one has, for the elastic form factors $B \to D^{(*)}$ \cite{FN}  :
\bea
\label{3.31e}
h_+(w) = \xi(w) + (\epsilon_c + \epsilon_b)L_1(w) + O^{h_+}_{1/{m_Q^2}}(w)
\eea
\bea
\label{3.32e}
h_-(w) = (\epsilon_c - \epsilon_b)L_4(w) + O^{h_-}_{1/{m_Q^2}}(w)
\eea
\bea
\label{3.33e}
h_V(w) = \xi(w) + \epsilon_c\left[ L_2(w) - L_5(w) \right] + \epsilon_b\left[ L_1(w) - L_4(w) \right] + O^{V}_{1/{m_Q^2}}(w)
\eea
\bea
\label{3.34e}
&&h_{A_1}(w) = \xi(w) + \epsilon_c\left[ L_2(w) - {w-1 \over w+1} L_5(w) \right] \nn \\ 
&&+\ \epsilon_b\left[ L_1(w) - {w-1 \over w+1} L_4(w) \right] + O^{A_1}_{1/{m_Q^2}}(w)
\eea
\bea
\label{3.35e}
h_{A_2}(w) = \epsilon_c\left[ L_3(w) + L_6(w) \right] + O^{A_2}_{1/{m_Q^2}}(w)
\eea
\bea
\label{3.36e}
&&h_{A_3}(w) = \xi(w) + \epsilon_c\left[L_2(w) - L_3(w) - L_5(w) + L_6(w) \right] \nn \\ 
&&+\ \epsilon_b\left[ L_1(w) - L_4(w) \right] + O^{A_3}_{1/{m_Q^2}}(w)
\eea

Luke's theorem \cite{L} states that, at first order in $1 \over m_Q$, one has
\bea
\label{3.37e}
L_1(1) = L_2(1) = 0
\eea

\noindent and therefore follows the important result that at zero recoil ($w = 1$) the subleading corrections to $h_+(1)$ and $h_{A_1}(1)$ begin at order $1/m_Q^2$ :
\bea
\label{3.38e}
h_+(1) = 1 + \delta^{h_+}_{1/{m_Q^2}} \qquad \qquad \qquad h_{A_1}(1) = 1 + \delta^{h_{A_1}}_{1/{m_Q^2}}
\eea

The functions $L_i(w)\ (i=4, 5, 6)$, corresponding to the so-called Current perturbations, are not independent according to HQET, and are given in terms of two independent functions $\overline{\Lambda} \xi(w)$ and $\xi_3(w)$ \cite{FN} :
\bea
\label{3.38-1e}
L_4(w) = -\overline{\Lambda} \xi(w) + 2\xi_3(w) \qquad \qquad \qquad \qquad
\eea
\bea
\label{3.38-2e}
L_5(w) = -\overline{\Lambda} \xi(w) \qquad \qquad \qquad \qquad \qquad \qquad
\eea
\bea
\label{3.38-3e}
L_6(w) = - {2 \over {w+1}} \left(\overline{\Lambda} \xi(w) + \xi_3(w)\right) \qquad \qquad
\eea

\noindent where $\xi(w)$ is the elastic IW function.\par 
One finds therefore the relation :
\bea
\label{3.38-4e}
L_4(w)+(1+w)L_6(w) = 3L_5(w) 
\eea

\noindent that reduces to the relation at zero recoil :
\bea
\label{3.38-5e}
L_4(1) + 2L_6(1) = 3L_5(1)
\eea

\subsubsection{Inelastic form factors $B \to D^{**}(0^+_{1/2}, 1^+_{1/2}, 1^+_{3/2}, 2^+_{3/2}) \ell \nu$ in HQET}

For the inelastic form factors $B \to D^{**}$ we reproduce only the leading order in the heavy quark expansion \cite{IW-2, LLSW}  :
\bea
\label{3.39e}
f_A(w) = - {w+1 \over \sqrt{2}}\ \tau_{3/2}(w) + O^{f_A}_{1/{m_Q}}(w)
\eea
\bea
\label{3.40e}
f_{V_1}(w) = {1-w^2 \over \sqrt{2}}\ \tau_{3/2}(w) + O^{f_{V_1}}_{1/{m_Q}}(w)
\eea
\bea
\label{3.41e}
f_{V_2}(w) = - {3 \over \sqrt{2}}\ \tau_{3/2}(w) + O^{f_{V_2}}_{1/{m_Q}}(w)
\eea
\bea
\label{3.42e}
f_{V_3}(w) = {w-2 \over \sqrt{2}}\ \tau_{3/2}(w) + O^{f_{V_3}}_{1/{m_Q}}(w)
\eea
\bea
\label{3.43e}
k_V(w) = - \sqrt{3}\tau_{3/2}(w) + O^{k_V}_{1/{m_Q}}(w)
\eea
\bea
\label{3.44e}
k_{A_1}(w) = - (w+1) \sqrt{3}\tau_{3/2}(w) + O^{k_{A_1}}_{1/{m_Q}}(w)
\eea
\bea
\label{3.45e}
k_{A_2}(w) = O^{k_{A_2}}_{1/{m_Q}}(w)
\eea
\bea
\label{3.46e}
k_{A_3}(w) = \sqrt{3}\tau_{3/2}(w) + O^{k_{A_3}}_{1/{m_Q}}(w)
\eea
\bea
\label{3.47e}
g_+(w) =  O^{g_+}_{1/{m_Q}}(w)
\eea
\bea
\label{3.48e}
g_-(w) = 2\tau_{1/2}(w) + O^{g_-}_{1/{m_Q}}(w)
\eea
\bea
\label{3.49e}
g_A(w) = 2\tau_{1/2}(w) + O^{g_A}_{1/{m_Q}}(w)
\eea
\bea
\label{3.50e}
g_{V_1}(w) = (w-1) 2\tau_{1/2}(w) + O^{g_{V_1}}_{1/{m_Q}}(w)
\eea
\bea
\label{3.51e}
g_{V_2}(w) = O^{g_{V_2}}_{1/{m_Q}}(w)
\eea
\bea
\label{3.52e}
g_{V_3}(w) = -2\tau_{1/2}(w) + O^{g_{V_3}}_{1/{m_Q}}(w)
\eea

\noindent where the different $O_{1/{m_Q}}(w)$ corrections are given in the detailed and careful paper by Leibovich et al. \cite{LLSW}. Among these corrections, we reproduce the ones that do not vanish at zero recoil, very relevant for what follows :
\bea
\label{3.53e}
g_+(1) = - 3(\epsilon_c+\epsilon_b)\Delta E_{1/2}\tau_{1/2}(1)
\eea
\bea
\label{3.54e}
g_{V_1}(1) = 2(\epsilon_c - 3\epsilon_b)\Delta E_{1/2}\tau_{1/2}(1) 
\eea
\bea
\label{3.55e}
f_{V_1}(1) = - 4\sqrt{2}\epsilon_c\Delta E_{3/2}\tau_{3/2}(1) 
\eea

\noindent where
\bea
\label{3.55bise}
\Delta E_{j} = m_{D(j^+)}-m_{D({1 \over 2}^-)} \qquad \qquad \qquad \left(j = {\scriptsize {1 \over 2},  {3 \over 2}}\right)  
\eea

\section{Bakamjian-Thomas approach to quark models}
\hspace*{\parindent}
As explained in \cite{LOPR-1}, the construction of the BT wave function in motion involves a {\it unitary} transformation that relates the wave function $\Psi_{s_1, \cdots ,s_n}^{(P)}({\vec{p}}_1, \cdots , {\vec{p}}_n)$ in terms of one-particle variables, the spin $\vec{ S}_i$ and momenta $\vec{p}_i$ to the so-called {\it internal} wave function $\Psi_{s_1, \cdots ,s_n}^{int}(\vec{P}, \vec{k}_2, \cdots ,\vec{k}_n)$ given in terms of another set of variables, the total momentum $\vec{P}$ and the internal momenta $ \vec{k}_1, \vec{k}_2, \cdots ,\vec{k}_n$ ($\sum\limits_i \vec{k}_i = 0$). This property ensures that, starting from an orthonormal set of internal wave functions, one gets an orthonormal set of wave functions in any frame. The base $\Psi_{s_1, \cdots , s_n}^{(P)}(\vec{p}_1, \cdots , \vec{p}_n)$ is useful to compute one-particle matrix elements like current one-quark matrix elements, while the second  $\Psi_{s_1, \cdots ,s_n}^{int}(\vec{P}, \vec{k}_2, \cdots ,\vec{k}_n)$ allows to exhibit Poincar\'e covariance. In order to satisfy the Poincar\'e commutators, the unique requirement is that the mass operator $M$, i.e. the Hamiltonian describing the spectrum at rest, should depend only on the internal variables and be rotational invariant, i.e. $M$ must commute with $\vec{P}$, ${\partial \over \partial \vec{P}}$ and $\vec{S}$. The internal wave function at rest $(2 \pi )^3 \delta (\vec{P}) \varphi_{s_1, \cdots , s_n}(\vec{k}_2, \cdots , \vec{k}_n)$ is an eigenstate of $M$, $\vec{P}$ (with $\vec{P} = 0$), $\vec{S}^2$ and $\vec{S}_z$,  while the wave function in motion of momentum $\vec{P}$ is obtained by applying the boost ${\bf B}_P$, where $P^0 = \sqrt{\vec{P}^2 + M^2}$ involves the dynamical operator $M$. \par

The final output of the formalism that gives the total wave function in motion $\Psi_{s_1, \cdots ,s_n}^{(P)}(\vec{p}_1, \cdots , \vec{p}_n)$ in terms of the internal wave function at rest $\varphi_{s_1, \cdots , s_n}(\vec{k}_2, \cdots , \vec{k}_n)$ is the formula
\bea
\label{3.1e}
&&\Psi_{s_1, \cdots ,s_n}^{(P)}(\vec{p}_1, \cdots , \vec{p}_n) = (2\pi )^3 \delta \left ( \sum_i \vec{p}_i - \vec{P}\right )  \sqrt{{\sum\limits_i p_i^0 \over M_0}}  \left ( \prod_i {\sqrt{k_i^0} \over \sqrt{p_i^0}}\right )\nn \\
&&\sum_{s'_1, \cdots , s'_n} \left [ D_i({\bf R}_i)\right ]_{s_i,s'_i} \varphi _{s'_1, \cdots , s'_n} (\vec{k}_2, \cdots , \vec{k}_n)
\eea

\noi where $p_i^0 = \sqrt{\vec{p}_i^2 + m_i^2}$ and $M_0$ is the free mass operator, given by 
\bea
\label{3.1-ae}
M_0 = \sqrt{(\sum\limits_i p_i)^2}
\eea

The internal momenta of the hadron at rest are given in terms of the momenta of the hadron in motion by the {\it free boost} 
\bea
\label{3.1-be}
k_i = {\bf B}_{\sum\limits_i p_i}^{-1} p_i
\eea

\noindent where the operator ${\bf B}_p$ is the boost $(\sqrt{p^2}, \vec{0}) \to p$, the Wigner rotations ${\bf R}_i$ in the preceding expression are
\bea
\label{3.1-ce}
{\bf R}_i = {\bf B}_{p_i}^{-1}{\bf B}_{\sum\limits_i p_i}^{-1} {\bf B}_{k_i} 
\eea

\noindent and the states are normalized by 
\bea
\label{3.1-de}
<\vec{P'}, {S'}_z| \vec{P}, {S}_z> \ = \ (2 \pi )^3 \delta ( \vec{P'} - \vec{P}) \delta_{S_z,S'_z}
\eea

The one-quark current  matrix element acting on quark 1 between two hadrons is then given by the expression
\bea
\label{3.2e}
&&< \vec{P'}, {S}'_z| J^{(1)}| \vec{P}, {S}_z >\ = \int {d \vec{p'}_1 \over (2 \pi)^3}\ {d \vec{p}_1 \over (2 \pi)^3} \left ( \prod_{i=2}^n {d \vec{p}_i \over (2 \pi)^3}\right )\nn \\
&&\Psi_{s'_1, \cdots ,s_n}^{P'} (\vec{p'}_1 , \cdots , \vec{p}_n)^* < \vec{p'}_1, {s'}_1|J^{(1)}|\vec{p}_1, {s}_1>\Psi_{s_1, \cdots , s_n}^{P} (\vec{p}_1 , \cdots ,\vec{p}_n)
\eea

\noi where $\Psi_{s_1, \cdots , s_n}^{P} (\vec{p}_1 , \cdots , \vec{p}_n)$ is given in terms of the internal wave function by (\ref{3.1e}) and $< \vec{p'}_1, s'_1|J^{(1)}|\vec{p}_1, s_1>$ is the one-quark current matrix element.\par

As demonstrated in \cite{LOPR-1,MLOPR-1}, in this formalism, in the heavy quark limit, current matrix elements are covariant and exhibit Isgur-Wise scaling, and one can compute Isgur-Wise functions like $\xi (w)$, $\tau_{1/2}(w)$, $\tau_{3/2}(w)$ \cite{MLOPR-3}. \par

After having presented the general calculations, there will remain to specify the mass operator $M$, which will be chosen as the one of the Godfrey and Isgur model in the following section.\par

We are interested in this paper in transitions between heavy quarks $b \to c$ where the initial meson is a pseudoscalar $\overline{B}$. We particularize the general formula (\ref{3.2e}) to the meson case $q_1\overline{q}_2$ where $q_1 \to q'_1$ labels the heavy quarks, $\overline{q}_2$ the light antiquark and the current operator $J^{(1)}$ acts on the heavy quark.\par 
As shown in \cite{LOPR-1}, one can express (\ref{3.2e}) in a Pauli matrix formalism and then in a Dirac matrix formalism. We reproduce here the needed master formula in the Dirac formalism :
\bea
\label{3.3e}
&&< \vec{P'}, \epsilon' | J^{(1)}| \vec{P}, \epsilon >\ = \int {d \vec{p}_2 \over (2 \pi)^3} {1 \over p^0_2}\ F(\vec{p}_2,\vec{P}',\vec{P})\nn \\
&& {1 \over 16}\ Tr \left [ O \left ( m_1 + {/ \hskip-2.5 truemm p}_1 \right ) (1 + {/ \hskip-2.5 truemm u}) \left ( m_2 + {/ \hskip-2.5 truemm p}_2 \right ) \Gamma_{u'} (1 + {/ \hskip-2. truemm u'}) \left ( m_1 + {/ \hskip-2. truemm p'}_1 \right )\right ]\ \varphi'(\vec{k}'_2)^* \varphi(\vec{k}_2) \qquad
\eea

\noindent where
\bea
\label{3.4e}
&&F(\vec{p}_2,\vec{P}',\vec{P})\ = {\sqrt{u^0u'^0} \over {p{^0_1} p'{^0_1}}}\sqrt{k{_1^0}k{_2^0} \over (k{_1^0}+m_1)(k{_2^0}+m_2)}\sqrt{k'{_1^0}k'{_2^0} \over (k'{_1^0}+m_1)(k'{_2^0}+m_2)}
\eea

\noindent In formula (\ref{3.3e}) the following unit four-vectors are used 
\bea
\label{3.5e}
u = {p_1+p_2 \over M_0} \qquad \qquad \qquad u' = {p'_1+p_2 \over M'_0}
\eea
\noindent with $M_0 = \sqrt{(p_1+p_2)^2}, M'_0 = \sqrt{(p'_1+p_2)^2}$, as explained above.\par 
In (\ref{3.3e}) the Dirac matrix $O$ depends on the current, for example $O = \gamma^\mu$ or $O = \gamma^\mu \gamma_5$ for the vector or axial current. On the other hand, the Dirac matrix $\Gamma_{u'}$ depends on the quantum numbers of the final state, namely $0^-, 1^-$ for the ground state and $0^+$, two $1^+$ states and $2^+$ for the excited states. Let us give now these matrices for the different $D$ states \cite{LOPR-1, MLOPR-2, M} :
\bea
\label{3.6e}
&&D(0^-) \qquad \qquad \qquad \ \ \ \  \Gamma_{u'} = 1\nn \\
&&D^*(1^-) \qquad \qquad \qquad \ \ \  \Gamma_{u'} = \gamma_5{/ \hskip-1.8 truemm \epsilon'^*_{u'}}\nn \\
&&D^{**}(0^+_{1/2}) \qquad \qquad \qquad \ \Gamma_{u'} = - {[{/ \hskip-1.8 truemm p_2}-(p_2.u'){/ \hskip-1.8 truemm u'} ] \gamma_5 \over \sqrt{(p_2.u')^2-m^2_2}}\nn \\
&&D^{**}(1^+_{1/2}) \qquad \qquad \qquad \Gamma_{u'} = - {[\epsilon^*_{u'}.p_2 + i \epsilon_{\alpha \beta \rho \sigma}u'^\alpha \epsilon^{*\beta}_{u'}p_2^\rho\gamma^\sigma\gamma_5] \over \sqrt{(p_2.u')^2-m^2_2}}\nn \\
&&D^{**}(1^+_{3/2}) \qquad \qquad \qquad \Gamma_{u'} = - {1 \over \sqrt{2}}{[2\epsilon^*_{u'}.p_2 - i \epsilon_{\alpha \beta \rho \sigma}u'^\alpha \epsilon^{*\beta}_{u'}p_2^\rho\gamma^\sigma\gamma_5] \over \sqrt{(p_2.u')^2-m^2_2}}\nn \\
&&D^{**}(2^+_{3/2}) \qquad \qquad \qquad \ \Gamma_{u'} = - \sqrt{3} {\gamma_\mu p_{2\nu} \epsilon^{* \mu\nu}_{u'\ } \gamma_5 \over \sqrt{(p_2.u')^2-m^2_2}}
\eea

\noindent where the convention $\epsilon_{0123} = -1$ is adopted, $\epsilon_{u'}$ are the polarizations relative to the four-vector $u'$, four-vectors for the $J^P = 1^P$ ($P = -, +$) states, and a tensor for the $J^P = 2^+$ state.

\subsection{Matrix elements in the heavy quark limit}

We now consider the heavy mass limit, defined as $m_1, m'_1 \to \infty$ with $v' = P'/M'$ and $v = P/M$ fixed, and $M/m_1 \to 1, M'/m'_1 \to 1$. One has, in this limit 
\bea
\label{3.7e}
&&{p_1 \over m_1} \to v, \qquad {p'_1 \over m'_1} \to v', \qquad \qquad {k_1^0 \over m_1} \to v, \qquad {k{'}_1^0 \over m'_1} \to 1 \nn \\
&&u \to v, \ \ u' \to v', \qquad \epsilon'_{u'} \to \epsilon'_{v'} = \epsilon', \qquad k_2 \to {\bf B}_v^{-1}p_2, \ \ k'_2 \to {\bf B}_{v'}^{-1}p_2
\qquad \eea

\noindent On the other hand, one has, due to the invariance of the scalar product,
\bea
\label{3.8e}
({\bf B}_v^{-1}p_2)^0 = p_2.v, \qquad ({\bf B}_{v'}^{-1}p_2)^0 = p_2.v'
\eea

\noindent and therefore the matrix element (\ref{3.3e}),(\ref{3.4e}) is given by the following covariant expression
\bea
\label{3.9e}
&&< \vec{P'}, \epsilon' | J^{(1)}| \vec{P}, \epsilon >\ = {1 \over \sqrt{4v^0v'^0}}\int  {d \vec{p}_2 \over (2 \pi)^3} {1 \over p^0_2}\ \sqrt{{(p_2.v')(p_2.v) \over (p_2.v'+m_2))(p_2.v+m_2))}}\nn \\
&& {1 \over 4}\ Tr\left [ O (1 + {/ \hskip-2.5 truemm v}) \left ( m_2 + {/ \hskip-2.5 truemm p}_2 \right ) \Gamma_{v'} (1 + {/ \hskip-2. truemm v'})\right ]\ \varphi'(\overrightarrow{{\bf B}_{v'}^{-1}p_2})^*\ \varphi(\overrightarrow{{\bf B}_{v}^{-1}p_2})
\eea
\noindent where the Dirac matrices $\Gamma_{v'}$ are identical to $\Gamma_{u'}$ in (\ref{3.6e}) with $u'$ replaced by the four-velocity $v'$.\par

The radial wave functions $\varphi'(\vec{k})$ and $\varphi(\vec{k})$ depend only on $\vec{k}^2$, and from (\ref{3.8e}) one has
\bea
\label{3.10e}
(\overrightarrow{{\bf B}_{v'}^{-1}p_2})^2 = (p_2.v')^2-m_2^2\ , \qquad \qquad (\overrightarrow{{\bf B}_{v}^{-1}p_2})^2 = (p_2.v)^2-m_2^2
\eea

\subsection {The Isgur-Wise functions $\xi(w)$, $\tau_{1/2}(w)$ and $\tau_{3/2}(w)$}

From the matrix elements (\ref{3.9e}), the operators (\ref{3.6e}) and the definitions and $1/m_Q$ expansion of the form factors given in Section 2, the Isgur-Wise functions $\xi(w), \tau_{1/2}(w),$ $\tau_{3/2}(w)$ are given by the expressions
\bea
\label{3.12e}
&&\xi(w)\ = {1 \over w+1}\int  {d \vec{p}_2 \over (2 \pi)^3} {1 \over p^0_2}\ \sqrt{{(p_2.v')(p_2.v) \over (p_2.v'+m_2)(p_2.v+m_2)}}\nn \\
&& [p_2.(v'+v)+m_2(w+1)] \ \varphi(\sqrt{(p_2.v')^2-m_2^2})^*\ \varphi(\sqrt{(p_2.v)^2-m_2^2})
\eea
\bea
\label{3.14e}
&& \tau_{1/2}(w) = {1 \over 2(1-w)}  \int  {d \vec{p}_2 \over (2 \pi)^3} {1 \over p^0_2}\ \sqrt{(p_2.v')(p_2.v) \over (p_2.v'+m_2)(p_2.v+m_2)}\nn \\
&& [(p_2.v)(p_2.v'+m_2)-(p_2.v')(p_2.v'+wm_2)+(1-w)m_2^2]\nn \\ 
&& {\varphi_{{1 \over 2}^+}(\sqrt{(p_2.v')^2-m_2^2})^*\ \varphi(\sqrt{(p_2.v)^2-m_2^2}) \over \sqrt{(p_2.v')^2-m_2^2}}
\eea
\bea
\label{3.15e}
&& \tau_{3/2}(w) = {1 \over 2\sqrt{3}(1-w)(1+w)^2}  \int  {d \vec{p}_2 \over (2 \pi)^3} {1 \over p^0_2}\ \sqrt{{(p_2.v')(p_2.v) \over (p_2.v'+m_2)(p_2.v+m_2)}}\nn \\
&& \{ 3[p_2.(v+v')]^2-2(w+1)(p_2.v)(2p_2.v'-m_2)-2(w+1)(p_2.v')(p_2.v'+wm_2)\nn \\ 
&& +(w^2-1)m_2^2 \}\ {\varphi_{{3 \over 2}^+}(\sqrt{(p_2.v')^2-m_2^2})^*\ \varphi(\sqrt{(p_2.v)^2-m_2^2}) \over \sqrt{(p_2.v')^2-m_2^2}}
\eea

\noindent where all the radial wave functions for the ${1 \over 2}^-, {1 \over 2}^+, {3 \over 2}^+$ states in the heavy quark limit are normalized by
\bea
\label{3.13e}
\int  {d \vec{p}_2 \over (2 \pi)^3} \mid\varphi(\vec{p}_2) \mid^2\ = 1
\eea   

\section{Limitations of the BT model at finite mass : choice of a convenient reference frame}

As we have emphasized above, the BT model provides a Poincar\' e covariant description of the states in motion, and also a Lorentz invariant formulation of the current matrix elements {\it in the heavy quark limit}. In the present paper we are interested in studying the $1/m_Q$ corrections to the matrix elements. However, although the current matrix elements can be formulated in the BT model by (\ref{3.3e}),(\ref{3.4e}), this expression is not Lorentz covariant.\par 
Another important point, also a limitation of the BT model, is that at finite mass, although one has lost Lorentz covariance, one does not even have Galilean covariance. In order to have Galilean covariance one would need to take the full non-relativistic limit, i.e. to consider the non-relativistic quark model : the model must be non-relativistic, not only for the heavy quarks $b$ and $c$, but also for the light quark.\par 
However, the non-relativistic quark model is not suited for our purpose, because what we want is to understand the {\it departures} relatively to the heavy quark limit predictions of the BT model due to the finiteness of the masses $m_b$ and $m_c$.\par 
Then, we are left to consider the BT model at finite mass in a definite reference frame. How to choose this frame ? Fortunately, there is a theoretical criterium for choosing a convenient frame. Namely, we will adopt the frame that is consistent with known theoretical results in the $1/m_Q$ expansion of HQET.\par
In Appendix D we have formulated a set of collinear frames, that go from the $B$ meson rest frame to the $D$ meson rest frame, dependent on a single parameter $\alpha$. The $B$ and $D$ rest frames correspond respectively to $\alpha = 0$ and $\alpha = 1$. There is an intermediate frame, that we call Equal Velocity Frame (EVF), in which the spatial velocities are equal in modulus ($v^0 = v'^0, v^z = -v'^z$), that corresponds to the value $\alpha = {1 \over 2}$. In this latter frame, the initial and final velocities then write, in terms of the variable $w = v.v'$ :

\bea
\label{3.13bise}
v = \left(\sqrt{{w+1 \over 2}}, 0, 0, - \sqrt{{w-1 \over 2}} \right) \qquad \qquad v' =  \left(\sqrt{{w+1 \over 2}}, 0, 0, \sqrt{{w-1 \over2}} \right) 
\eea

Considering the matrix element at arbitrary masses (\ref{3.3e}) for the ground state $B \to D^{(*)} \ell \nu$ transitions, and making analytically an expansion up to the first power in $1/m_c$ and $1/m_b$, we have realized that the form of the HQET expansion of the form factors as written in (\ref{3.31e})-(\ref{3.36e}) is not fulfilled in any of the considered collinear frames, {\it except in the EVF}. In this frame, relations (\ref{3.31e})-(\ref{3.36e}), at least up to first order in $1/m_Q$, are exactly satisfied. This seems to us a good enough criterium for choosing the EVF in our calculations. We will below compute all the ground state subleading functions $L_i(w) (i = 1,... 6)$ and verify also that Luke's theorem is satisfied.\par
A last important point of principle is in order here. Had we adopted the non-relativistic quark model (including the light quark), relations (\ref{3.31e})-(\ref{3.36e}) are exactly satisfied in any Galilean frame. However, as pointed out above, we need to consider the $b$ and $c$ quarks as heavy, and the spectator light quark as relativistic. Quantitatively, the results of the non-relativistic quark model would not make much sense in order to consider departures of the heavy quark limit results of the BT model due to the $b$ and $c$ finite masses.

\section{$1/m_Q$ form factors for the ground state transitions $B \to D^{(*)} \ell \nu$ in the BT model}

To make explicit the discussion of the $1/m_Q$ corrections to $B \to D^{(*)} \ell \nu$, let us rewrite the basic formulas at finite mass (\ref{3.3e}),(\ref{3.4e}) under the form and new notation 
\bea
\label{6.1e}
< D^{(*)}(\vec{P'}), \epsilon' | J^{(1)}| B(\vec{P}) >\ = \int {d \vec{p}_2 \over (2 \pi)^3} {1 \over p^0_2}\ G_{D^{(*)}B}(\vec{p}_2,\vec{P}',\vec{P})
\ \varphi'_{D^{(*)}}(\vec{k}'_2)^* \varphi_B(\vec{k}_2)
\eea
\noindent with
\bea
\label{6.2e}
&&G_{D^{(*)}}(\vec{p}_2,\vec{P}',\vec{P})\ = {\sqrt{u^0u'^0} \over {p{^0_1} p'{^0_1}}}\sqrt{k{_1^0}k{_2^0} \over (k{_1^0}+m_1)(k{_2^0}+m_2)}\sqrt{k'{_1^0}k'{_2^0} \over (k'{_1^0}+m_1)(k'{_2^0}+m_2)} \nn \\
&& {1 \over 16}\ Tr \left [ O \left ( m_1 + {/ \hskip-2.5 truemm p}_1 \right ) (1 + {/ \hskip-2.5 truemm u}) \left ( m_2 + {/ \hskip-2.5 truemm p}_2 \right ) \Gamma_{u'}^{D^{(*)}} (1 + {/ \hskip-2. truemm u'}) \left ( m_1 + {/ \hskip-2. truemm p'}_1 \right )\right ]\ 
\eea

\noindent where $\Gamma_{u'}^D = 1$ and $\Gamma_{u'}^{D^{*}} = \gamma_5{/ \hskip-1.8 truemm \epsilon'^*_{u'}}$.\par
For the sake of clarity we now adopt  the notation
\bea
\label{6.3e}
\epsilon_b = {1 \over {2m_1}} = {1 \over {2m_b}} \qquad \qquad \qquad \epsilon_c = {1 \over {2m'_1}} = {1 \over {2m_c} }
\eea

To compute the $1/{2m_Q}$ subleading functions $L_i(w)\ (i = 1,...6)$  (\ref{3.31e})-(\ref{3.36e}), we need to expand the matrix element (\ref{6.1e}) in powers of $\epsilon_b, \epsilon_c$ up to the first order. Simbolically we can write, simplifying the notation,
$$< D^{(*)}(\vec{P'}), \epsilon' | J^{(1)}| B(\vec{P}) >\ =\ < D^{(*)}(\vec{P'}), \epsilon' | J^{(1)}| B(\vec{P}) >_0$$ 
$$+ \int {d \vec{p}_2 \over (2 \pi)^3} {1 \over p^0_2}\ \left[ \epsilon_b G^{(b)}_0(\vec{p}_2,\vec{P}',\vec{P}) + \epsilon_c G^{(c)}_0(\vec{p}_2,\vec{P}',\vec{P}) \right]
\ \varphi'_0(\vec{k}'_2)^* \varphi_0(\vec{k}_2)$$
\bea
\label{6.4e}
+ \int {d \vec{p}_2 \over (2 \pi)^3} {1 \over p^0_2}\ G_0(\vec{p}_2,\vec{P}',\vec{P})
\ \left[ \epsilon_b \varphi'_0(\vec{k}'_2)^* \varphi_0^{(b)}(\vec{k}_2) + \epsilon_c \varphi{'}_0^{(c)}(\vec{k}'_2)^* \varphi_0(\vec{k}_2)\right] 
\eea

\noindent In the preceding equation, the subindex $0$ means $\epsilon_b = \epsilon_c = 0$ (heavy quark limit).\par 
We have separated the perturbation of the kernel $G$ and of the wave functions $\varphi$, in an obvious notation.
In what follows we will neglect the second term in (\ref{6.4e}) since we have realized numerically that the perturbation of the wave functions gives a very small contribution.\par
Using (\ref{6.4e}), it is convenient to write the matrix elements (\ref{3.31e})-(\ref{3.36e}) using the following notation :
\bea
\label{6.5e}
h_+(w) = \xi(w) + \epsilon_c H_+^{(c)}(w) + \epsilon_b H_+^{(b)}(w) + O^{h_+}_{1/{m_Q^2}}(w)
\eea
\bea
\label{6.6e}
h_-(w) = \epsilon_c H_-^{(c)}(w) + \epsilon_b H_-^{(b)}(w) + O^{h_-}_{1/{m_Q^2}}(w)
\eea
\bea
\label{6.7e}
h_V(w) = \xi(w) + \epsilon_c H_V^{(c)}(w) + \epsilon_b H_V^{(b)}(w) + O^{V}_{1/{m_Q^2}}(w)
\eea
\bea
\label{6.8e}
h_{A_1}(w) = \xi(w) +  \epsilon_c H_{A_1}^{(c)}(w) + \epsilon_b H_{A_1}^{(b)}(w) + O^{A_1}_{1/{m_Q^2}}(w)
\eea
\bea
\label{6.9e}
h_{A_2}(w) = \epsilon_c H_{A_2}^{(c)}(w) + \epsilon_b H_{A_2}^{(b)}(w) + O^{A_2}_{1/{m_Q^2}}(w)
\eea
\bea
\label{6.10e}
&&h_{A_3}(w) = \xi(w) + \epsilon_c H_{A_3}^{(c)}(w) + \epsilon_b H_{A_3}^{(b)}(w) + O^{A_3}_{1/{m_Q^2}}(w)
\eea

Performing analytically an expansion of the matrix elements for the different currents in powers of $\epsilon_b, \epsilon_c$, we can compute the different functions $H^{(Q)} (Q = b, c)$, and from them obtain the subleading functions $L_i(w)\ (i = 1,...6)$ appearing in (\ref{3.31e})-(\ref{3.36e}), by using the straightforward relations :
\bea
\label{6.11e}
L_1(w) = H_+^{(c)}(w) = H_+^{(b)}(w) = {1 \over 2} \left[ (w+1)H_{A_1}^{(b)}(w) - (w-1)H_{A_3}^{(b)}(w) \right]
\eea
\bea
\label{6.12e}
L_2(w) = {1 \over 2} \left[ (w+1)H_{A_1}^{(c)}(w) - (w-1)H_V^{(c)}(w) \right]
\eea
\bea
\label{6.13e}
L_3(w) = {1 \over 2} \left[ H_{A_2}^{(c)}(w) - H_{A_3}^{(c)}(w) + H_V^{(c)}(w) \right]
\eea
\bea
\label{6.14e}
L_4(w) = H_-^{(c)}(w) = -H_-^{(b)}(w) = {w+1 \over 2} \left[ H_{A_1}^{(b)}(w) - H_{A_3}^{(b)}(w) \right]
\eea
\bea
\label{6.15e}
L_5(w) = {w+1 \over 2} \left[ H_{A_1}^{(c)}(w) - H_V^{(c)}(w) \right]
\eea
\bea
\label{6.16e}
L_6(w) = {1 \over 2} \left[ H_{A_2}^{(c)}(w) + H_{A_3}^{(c)}(w) - H_V^{(c)}(w) \right]
\eea

From these relations, and the expressions for the different functions $H^{(Q)} (Q = b, c)$, we find analytically that Luke's theorem \cite{L} (\ref{3.37e}) is satisfied
\bea
\label{6.17e}
L_1(1) = L_2(1) = 0
\eea

Moreover we find, for the functions $L_i(w)\ (i=1, 2, 3)$, corresponding to the so-called Lagrangian perturbations, the following results, that do not follow from HQET, and are specific to the BT model :
\bea
\label{6.18e}
L_1(w) = L_2(w)\ , \qquad \qquad \qquad L_3(w) = 0 
\eea

In the BT model, for the functions $L_i(w)\ (i=4, 5, 6)$ that correspond to the Current perturbations, we find analytically relation (\ref{3.38-5e}) that holds in HQET :
\bea
\label{6.19e}
L_4(1) + 2L_6(1) = 3L_5(1)
\eea

More explicitly, we find in the limit $m_D = m_{D^*} = m_c + \overline{\Lambda}$, calling from now on the light quark mass $m_2 = m$ : 
\bea
\label{5.4-1e}
L_4(1) = - \overline{\Lambda} + {2 \over 3} \int  {d \vec{p} \over (2 \pi)^3}\ {\vec{p}^2 \over m + \sqrt{m^2+\vec{p}^2}}\ | \varphi(\vec{p})|^2
\eea
\bea
\label{5.4-2e}
L_5(1) = - \overline{\Lambda} \qquad \qquad \qquad \qquad \qquad \qquad \qquad \qquad 
\eea
\bea
\label{5.4-3e}
L_6(1) = - \overline{\Lambda} - {1 \over 3} \int  {d \vec{p} \over (2 \pi)^3}\ {\vec{p}^2 \over m + \sqrt{m^2+\vec{p}^2}}\ | \varphi(\vec{p})|^2
\eea

\noindent where the internal wave function normalization 
\bea
\label{5.4-4e}
\int  {d \vec{p} \over (2 \pi)^3}\ |\varphi(\vec{p})|^2 = 1
\eea

\noindent has been used. Relation (\ref{5.4-2e}) is in agreement with (\ref{3.38-2e}) at zero recoil.\par

\section{$1/m_Q$ form factors at zero recoil for transitions to excited states $B \to D^{**} \ell \nu$ in the BT model}

Performing a series expansion of the relevant form factors one finds, in the BT model, at zero recoil :
\bea
\label{7.1e}
g_+(1) = - 3 (\epsilon_c+\epsilon_b)\ {1 \over 3} \int  {d \vec{p} \over (2 \pi)^3}\ |\vec{p}|\ \varphi_{{1 \over 2}^+}(|\vec{p}|)^*\varphi(|\vec{p}|)
\eea
\bea
\label{7.2e}
g_{V_1}(1) = 2(\epsilon_c-3\epsilon_b)\ {1 \over 3}  \int  {d \vec{p} \over (2 \pi)^3}\ |\vec{p}|\ \varphi_{{1 \over 2}^+}(|\vec{p}|)^*\varphi(|\vec{p}|)
\eea
\bea
\label{7.3e}
f_{V_1}(1) = -4 \sqrt{2}\ \epsilon_c\ {1 \over 3} \int  {d \vec{p} \over (2 \pi)^3}\ |\vec{p}_2|\ \varphi_{{3 \over 2}^+}(|\vec{p}|)^*\varphi(|\vec{p}|)
\eea

\noindent These formulas hold for all collinear reference frames considered in Appendix D. We observe that the $1/m_Q$ dependence agrees with the prediction of HQET for all three form factors $g_+(1)$, $g_{V_1}(1)$ and $f_{V_1}(1)$ (formulas (\ref{3.53e})-(\ref{3.55e})), in particular the BT model predicts for the two states belonging to the same doublet $0^- \to 0^+_{1/2}$,  $0^- \to 1^+_{1/2}$ :
\bea
\label{7.4e}
{g_+(1) \over g_{V_1}(1)}  = -{3 \over 2} {\epsilon_c+\epsilon_b \over \epsilon_c-3\epsilon_b}
\eea

\noindent while the form factor $f_{V_1}(1)$ for $0^- \to 1^+_{3/2}$ is independent because a different radial wave function $\varphi_{{3 \over 2}^+}(|\vec{p}|)$ appears in formula (\ref{7.3e}). Formula (\ref{7.4e}) is consistent with the expectations of HQET (\ref{3.53e})-(\ref{3.55e}).\par

Another matter is the absolute magnitude of the BT results (\ref{7.1e})-(\ref{7.3e}) as compared with the HQET results by Leibovich et al. \cite{LLSW} (\ref{3.53e})-(\ref{3.55e}). In the latter expressions we see that there is factorization between the level spacings and the corresponding inelastic IW functions at zero recoil : $\Delta E_{{1 \over 2}}\tau_{1/2}(1)$ or $\Delta E_{{1 \over 2}}\tau_{3/2}(1)$.\par
The spin-orbit term is small and one can therefore assume that the level spacing is about the same for both $j^+$ states :  
\bea
\label{7.5e}
\Delta E_{{1 \over 2}} \simeq \Delta E_{{3 \over 2}}
\eea

Then, the form factors at zero recoil (\ref{3.53e})-(\ref{3.55e}) are in the ratios
\bea
\label{7.6e}
g_+(1) : g_{V_1}(1) : f_{V_1}(1) = - 3 (\epsilon_c+\epsilon_b) \tau_{1/2}(1) : 2(\epsilon_c-3\epsilon_b) \tau_{1/2}(1) : -4 \sqrt{2}\ \epsilon_c \tau_{3/2}(1)
\eea

\noindent while we find, from (\ref{7.1e})-(\ref{7.3e}), in the BT model within the same assumption of small spin-orbit coupling :
\bea
\label{7.7-1e}
g_+(1) : g_{V_1}(1) : f_{V_1}(1) = - 3 (\epsilon_c+\epsilon_b) : 2(\epsilon_c-3\epsilon_b) : -4 \sqrt{2}\ \epsilon_c
\eea

The contradiction between the results of HQET (\ref{7.6e}) and the ones of the BT model (\ref{7.7-1e}) is obvious because of the values (\ref{1.2e}) found in the heavy quark limit in the BT model (for the IG potential) : $\tau_{1/2}(1) = 0.22, \tau_{3/2}(1) = 0.54$.\par
The origin of the difference between $\tau_{1/2}(1)$ and $\tau_{3/2}(1)$ in the BT model is the following. From expressions (\ref{3.14e}),(\ref{3.15e}) one obtains at zero recoil \cite{MLOPR-1,M}
\bea
\label{7.7-2e}
\tau_{1/2}(1) = -m\ {1 \over 12 \pi^2} \int_0^\infty p^2 dp\ \varphi_{1/2}(p)\ \left[{p \over m+p^0} \left(3 + {m \over p^0} \right) + 2 {d \over dp}\right]\ \varphi(p)
\eea
\bea
\label{7.7-3e}
\tau_{3/2}(1) = -m\ {1 \over 12 \pi^2} \int_0^\infty p^2 dp\ \varphi_{3/2}(p)\ \left[{p \over m+p^0} {m \over p^0} + 2 {d \over dp}\right]\ \varphi(p) \qquad \ \ \  
\eea

Therefore, due to the first terms in the r.h.s. of (\ref{7.7-2e}) and (\ref{7.7-3e}) one gets in the BT model $\tau_{1/2}(1) \not = \tau_{3/2}(1)$. As analyzed in detail in \cite{LOPR-2} the Wigner rotations are at the origin of these terms : 
\bea
\label{7.7-4e}
\tau_j(1) \sim \ \left< j^+\left|- {p^0iz+izp^0 \over 2} + {i \over 2} {(\vec{\sigma}\times\vec{p}_T)_z \over p^0+m} \right|{1 \over 2}^-\right> \qquad \qquad \left(j = {1 \over 2}, {3 \over 2} \right) 
\eea

\noindent The Wigner rotation, second term in (\ref{7.7-4e}) is a relativistic effect dependent on the spin that gives the difference between $\tau_{1/2}(1)$ and $\tau_{3/2}(1)$.\par

\subsection{BT model $1/m_Q$ form factors at zero recoil for transitions to excited states in the non-relativistic limit}

Let us first observe that expressions (\ref{7.1e})-(\ref{7.3e}) are independent of the light quark mass $m$. Therefore, the same expressions must be valid {\it in the non-relativistic limit of the BT model}, i.e. taking $|\vec{p}| << m$. Let us assume this limit and consider the non-relativistic Hamiltonian for the light quark interacting with the heavy quark :
\bea
\label{7.7e}
H = {p^2 \over 2m} + V(r)
\eea

\noindent  where $\vec{r}$ is the relative position between the light quark and the heavy quark.\par

Let us first remark that in the non-relativistic limit, since the spin-orbit term does not contribute, one has 
\bea
\label{7.8e}
\varphi_{{1 \over 2}}(p) = \varphi_{{3 \over 2}}(p)\ , \qquad \qquad \Delta E_{{1 \over 2}} = \Delta E_{{3 \over 2}}  
\eea

\noindent In the non-relativistic limit one has $\tau_{1/2}(w) = \tau_{3/2}(w)$, that at zero recoil is given by
\bea
\label{7.9e}
\tau_j(1) = -m\ {1 \over 6 \pi^2} \int_0^\infty p^2 dp\  \varphi_j(p)\ {d \over dp}\ \varphi(p) = - {1 \over 3}\ m \left< 0^+ \left| {d \over dp} \right| 0^-\right> \ \ \  \left( j = {1 \over 2}, {3 \over 2} \right)
\eea

Using (\ref{7.9e}) and the non-relativistic Hamiltonian (\ref{7.7e}) let us compute 
$$\Delta E_j \tau_j(1) = - {1 \over 3}\ m \left< 0^+ \left| \left[H, {d \over dp}\right] \right| 0^-\right>
= - {1 \over 3}\ m \left< 0^+ \left| \left[{p^2 \over 2m}, {d \over dp}\right] \right| 0^-\right>$$
\bea
\label{7.10e}
= {1 \over 6 \pi^2} \int_0^\infty p^3 dp\ \varphi_j(p) \varphi(p) \qquad \qquad \qquad \left( j = {{1 \over 2}, {3 \over 2}} \right)
\eea

\noindent and we obtain therefore the common factor in the r.h.s. of  eqns. (\ref{7.1e})-(\ref{7.3e}).\par
Finally, in the non-relativistic limit we obtain relations (\ref{3.53e})-(\ref{3.55e}) with $\Delta E_{{1 \over 2}} \tau_{{1 \over 2}}(1) = \Delta E_{{3 \over 2}} \tau_{{3 \over 2}}(1)$ given by the r.h.s. of (\ref{7.10e}).\par

The argument has a transparent physical interpretation in configuration space. In the non-relativistic limit of (\ref{7.7-4e}) the Wigner rotations are subleading and one has
\bea
\label{7.11e}
\tau_j(1) \sim \ m \left< j^+ \left| -iz \right| {1 \over 2}^-\right> \qquad \qquad \left(j = {1 \over 2}, {3 \over 2} \right)
\eea

\noindent Computing the matrix element of the axial current $A^0$ at zero recoil one has, since the active quark is the heavy quark labelled 1 :
\bea
\label{7.12e}
<0^+|A^0|0^->\ \sim \left( {1 \over m_c} + {1 \over m_b} \right) < 0^+ | p_{1z} | 0^-> \qquad \qquad \left(j = {1 \over 2}, {3 \over 2} \right)
\eea

\noindent then one has, from the non-relativistic Hamiltonian (\ref{7.7e}) and $\vec{p}_1 = -\vec{p}_2 = -\vec{p}$, where $\vec{p}$ is the momentum of the light spectator quark :
$$< 0^+ | p_{1z} | 0^->\ = m \left< 0^+ \left| - {p_z \over m} \right| 0^-\right>$$
\bea
\label{7.12e}
= - i m < 0^+| [H,z] | 0^- > = -i m (E_1-E_0) < 0^+| z | 0^- >
\eea

\noindent where $E_0, E_1$ are the energies of the ground state and the excited state. Therefore, the dependence on the level spacing of HQET follows in the non-relativistic limit, as we have already seen from (\ref{7.10e}). 
 
\section{The Godfrey-Isgur quark model for spectroscopy}

Let us now particularize the above expressions for the choice of the mass operator $M$ given by the Godfrey-Isgur model \cite{GI}, and perform the numerical calculations.\par
The GI model for meson spectroscopy describes the whole set of meson spectra $q\overline{q}$ and $Q\overline{q}$, where $q$ is a light quark ($q$ = $u$, $d$, $s$) and $Q$ is a heavy quark ($Q$ = $c$, $b$), with the important exception of the recently discovered narrow states $D_{sJ}$ ($0^+$ and $1^+$), that are too low in mass compared with the predictions of the model. The model contains a relativistic kinetic term of the form
\bea
\label{4.1e}
K = \sqrt{\vec{k}^2_1+m{^2_1}} + \sqrt{\vec{k}^2_2+m{^2_2}}
\eea
that is identical to the operator $M_0$ at rest, and a complicated interaction term that includes : (1) a Coulomb part with a $q^2$ dependent $\alpha_s$, (2) a linear confining piece, and (3) terms describing the spin-orbit and spin-spin interactions. All singularities are regularized - e.g. terms of the type $\delta(\vec{r})$ or $1/m_2$, where $m_2$ is the light quark mass. The hamiltonian H depends on a number of parameters that are fitted to describe all the meson spectra.

\section {Form factors for the ground state $B \to D^{(*)} \ell \nu$}

This Section contains the numerical results for the ground state form factors $B \to D^{(*)} \ell \nu$ using the Bakamjian-Thomas model exposed above and the internal wave functions provided by the GI spectrocopic potential, given in Appendices B (heavy quark limit) and C (at finite mass).\par
In Fig. 1 we give the prediction for the elastic IW function $\xi(w)$ and in Figs. 2-7 we give the results for the different $B \to D^{(*)} \ell \nu$ form factors at finite mass compared with their heavy quark limit. The finite mass effect is rather small in general, even in the case of the form factors that vanish in the heavy quark limit, $h_-(w)$ and $h_{A_2}(w)$.

\includegraphics[scale=1.]{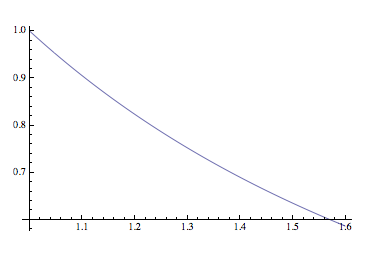}\\ 
Fig. 1. The elastic Isgur-Wise function $\xi(w) = \left({2 \over w+1}\right)^{2\rho^2}$ in the BT model ($\rho^2 = 1.023$).

\includegraphics[scale=1.]{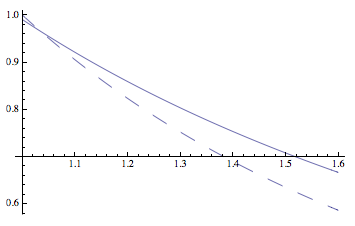}\\
Fig. 2. The form factor $h_+(w)$ in the BT model at finite mass (continuous line, $h_+(1) = 0.99033$) and in the heavy quark limit (dashed line).

\includegraphics[scale=1.]{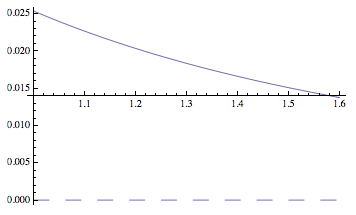}\\ 
Fig. 3. The form factor $h_-(w)$ in the BT model at finite mass (continuous line, $h_-(1) = 0.02535$) and in the heavy quark limit (dashed line).

\vskip 0.5 truecm

\includegraphics[scale=1.]{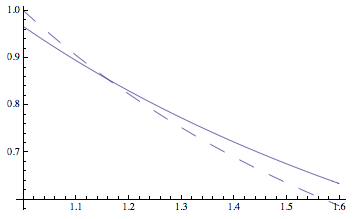}\\ 
Fig. 4. The form factor $h_{A_1}(w)$ in the BT model at finite mass (continuous line, $h_{A_1}(1) = 0.96606$) and in the heavy quark limit (dashed line).

\includegraphics[scale=1.]{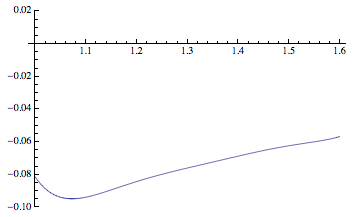}\\ 
Fig. 5. The form factor $h_{A_2}(w)$ at finite mass in the BT model (it vanishes at infinite mass).

\vskip 0.5 truecm

\includegraphics[scale=1.]{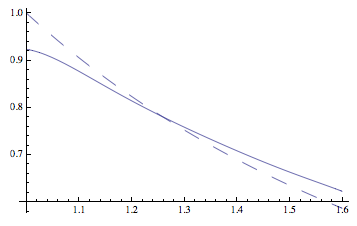}\\ 
Fig. 6. The form factor $h_{A_3}(w)$ in the BT model at finite mass (continuous line, $h_{A_3}(1) = 0.92299$)  and in the heavy quark limit (dashed line).

\includegraphics[scale=1.]{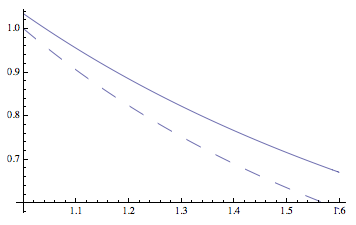}\\ 
Fig. 7. The form factor $h_V(w)$ in the BT model at finite mass (continuous line, $h_V(1) = 1.03414$)  and in the heavy quark limit (dashed line).

\subsection {First order $1/m_Q$ functions and Luke theorem}

Here we compute within the BT model with the GI internal wave functions the subleading functions $L_i(w)$ defined in (\ref{3.31e})-(\ref{3.36e}) and given by equations (\ref{6.11e})-(\ref{6.16e}) in terms of the functions $H^{(Q)}$. In the results given below we consider only the expansion of the kernel $G$ in (\ref{6.4e}), since the perturbation of the wave function $\varphi$ gives a negligible numerical contribution.

\includegraphics[scale=1.]{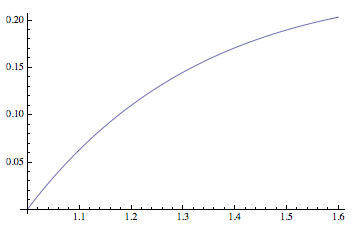}\\ 
Fig. 8. The subleading functions $L_1(w), L_2(w)$ in the BT model, in which $L_1(w) = L_2(w)$ (in GeV units). Luke's theorem $L_1(1) = L_2(1) = 0$ is satisfied.

\vskip 0.5 truecm

Notice that for the other elastic Lagrangian perturbation $L_3(w)$ in the BT model we find $L_3(w) = 0$, eqn. (\ref{6.18e}).

\vskip 0.5 truecm

\includegraphics[scale=1.]{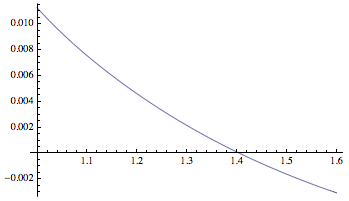}\\ 
Fig. 9. The subleading function $L_4(w)$ in the BT model ($L_4(1) = 0.011250\ \rm{GeV}$).

\vskip 0.5 truecm

\includegraphics[scale=1.]{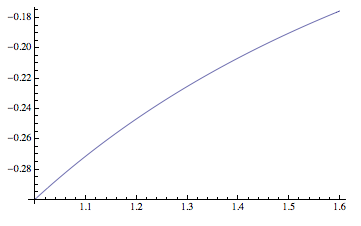}\\ 
Fig. 10. The subleading function $L_5(w)$ in the BT model ($L_5(1) = -\overline{\Lambda} = -0.3\ \rm{GeV}$).

\vskip 0.5 truecm

\includegraphics[scale=1.]{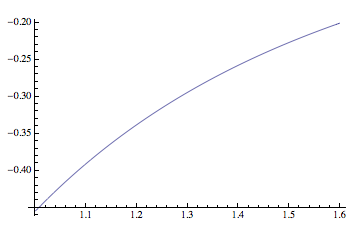}\\ 
Fig. 11. The subleading function $L_6(w)$ in the BT model ($L_6(1) = -0.455474\ \rm{GeV}$).

\vskip 0.5 truecm

Some comments are in order concerning these figures.\par
Let us begin with the Lagrangian perturbation functions $L_i(w)\ (i=1, 2, 3)$. 
First, we observe that Luke's theorem \cite{L} (\ref{3.37e}) is indeed satisfied :
\bea
\label{5.1e}
L_1(1) = L_2(1) = 0
\eea

On the other hand, the result that we find for $L_3(w)$, $L_3(1) = 0$, is not a prediction of HQET.\par
Considering now the Current perturbation functions $L_i(w)\ (i=4, 5, 6)$, these functions are not independent according to HQET, and are given in terms of two independent functions $\overline{\Lambda} \xi(w)$ and $\xi_3(w)$ \cite{FN} (\ref{3.38-1e})-(\ref{3.38-3e}). We recall here the expression of $L_5(w)$ in terms of the elastic IW function $\xi(w)$ :
\bea
\label{5.2e}
L_5(w) = -\overline{\Lambda} \xi(w) \qquad \qquad \qquad \qquad
\eea

\noindent and the linear relation 
\bea
\label{5.3e}
L_4(w)+(1+w)L_6(w) = 3L_5(w) 
\eea 

It is important to emphasize that relation (\ref{5.2e}) is in analytical agreement with the prediction of the BT model for the elastic IW function (Fig. 1, where $\overline{\Lambda} = 0.3$ GeV). From the explicit formulae for $L_4(w), L_5(w)$, and $L_6(w)$ in the BT model, we have checked that this relation is also analytically exact within the model.\par
From this section we conclude that the BT model gives a description of the corrections of $O(1/m_Q)$ to the elastic form factors that is consistent with the predictions of HQET, even for their $w$-dependence. 

\subsection {$1/m_Q^2$ corrections at zero recoil for $h_+(w)$ and $h_{A_1}(w)$}

In the BT model we find indeed that the results satisfy Luke's theorem (\ref{3.37e}), and therefore the corrections at zero recoil to $h_+(1)$ and $h_{A_1}(1)$ begin at order $1/m_Q^2$, eqn. (\ref{3.38e}). We get for the sum of all orders $1/{m_Q^n}\ (n \geq 2)$ that contribute at zero recoil 
\bea
\label{5.4e}
- \sum_{n \geq 2} \delta^{h_+}_{1/{m_Q^n}} = 0.0097
\eea
\bea
\label{5.5e}
- \sum_{n \geq 2} \delta^{h_{A_1}}_{1/{m_Q^n}} = 0.0339
\eea

These results can be compared with the $O(1/{m_Q^2})$ power corrections obtained in HQET \cite{BSUV}. To do that we must switch off the hard gluon radiative corrections in the HQET approach. For the current masses $m_c = 1.25\ \rm{GeV}$, $m_b = 4.75\ \rm{GeV}$ and $\mu_{G}^2 = 0.35\ \rm{GeV}^2$, $\mu_{\pi}^2 = 0.40\ \rm{GeV}^2$, the second order HQET power corrections are roughly $- \delta^{h_+}_{1/{m_Q^2}} \simeq 0.0022, - \delta^{h_{A_1}}_{1/{m_Q^2}} \simeq 0.042$, to be compared with the precedent results of the BT model for the power corrections to all orders with the constituent masses of the model.  

\section {Form factors for the excited states $B \to D^{**} \ell \nu$}

This Section contains the numerical results for the inelastic form factors $B \to D^{(**)} \ell \nu$, using the BT model and the internal wave functions provided by the GI potential tabulated in Appendix B (heavy quark limit) and Appendix C (finite mass).\par
In Figs. 12 and 13 we give the predictions for the inelastic IW functions $\tau_{1/2}(w)$ and $\tau_{3/2}(w)$.

\vskip 0.5 truecm

\includegraphics[scale=1.]{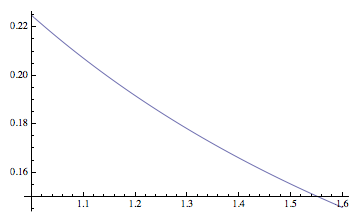}\\ 
Fig. 12. The IW function $\tau_{1/2}(w) = \tau_{1/2}(1) \left( {2 \over {w+1}} \right)^{2\sigma^2_{1/2}}$ for the transitions $0^- \to 0^+_{1/2}, 1^+_{1/2}$ in the BT model with the GI hamiltonian ($\tau_{1/2}(1) = 0.2248$, $\sigma^2_{1/2} = 0.84$).

\vskip 0.5 truecm

\includegraphics[scale=1.]{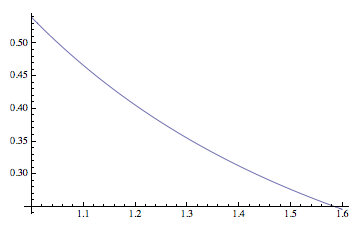}\\ 
Fig. 13. The IW function  $\tau_{3/2}(w) = \tau_{3/2}(1) \left( {2 \over {w+1}} \right)^{2\sigma^2_{3/2}}$ for the transitions $0^- \to 1^+_{3/2}, 2^+_{3/2}$ in the BT model with the GI hamiltonian ($\tau_{3/2}(1) = 0.5394$, $\sigma^2_{3/2} = 1.50$).

\vskip 1.0 truecm

In Figs. 14-27 we give the results for the different form factors contributing to the transitions $B \to D^{(**)} ( 0^+_{1/2},  1^+_{1/2},  1^+_{3/2},  2^+_{3/2})$. In the figures we compare the results at finite mass with the corresponding heavy quark limit.\par 
Unlike the elastic case, the finite mass effects for these inelastic form factors are not small, even for some form factors that vanish in the heavy quark limit. This is particularly true for the transition $0^- \to 0^+$. In this case, the leading form factor $g_-(w)$ is reduced by about a factor 1.5, while the absolute magnitude of the form factor $g_+(w)$, that vanishes in the heavy quark limit, becomes of the same order as the leading one.  

\includegraphics[scale=1.]{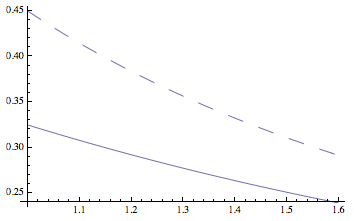}\\ 
Fig. 14. The form factor  $g_-(w)$ for the transition $0^- \to 0^+$ in the BT model at finite mass (full line, $g_-(1) = 0.3241$) and in the heavy quark limit (dashed line).

\vskip 0.5 truecm

\includegraphics[scale=1.]{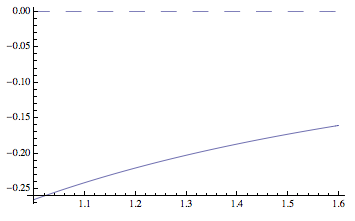}\\ 
Fig. 15. The form factor  $g_+(w)$ for the transition $0^- \to 0^+$ in the BT model at finite mass ($g_+(1) = -0.2657$) and in the heavy quark limit (dashed line).

\vskip 0.5 truecm

\includegraphics[scale=1.]{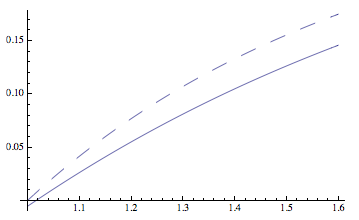}\\
Fig. 16.  $g_{V_1}(w)$ for the transition $0^- \to 1^+_{1/2}$ in the BT model at finite mass ($g_{V_1}(1) = -0.0022$) and in the heavy quark limit (dashed line).

\vskip 0.5 truecm

\includegraphics[scale=1.]{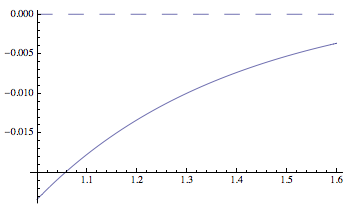}\\ 
Fig. 17.  $g_{V_2}(w)$ for the transition $0^- \to 1^+_{1/2}$ in the BT model at finite mass ($g_{V_2}(1) = -0.0159$) and in the heavy quark limit (dashed line).

\vskip 0.5 truecm

\includegraphics[scale=1.]{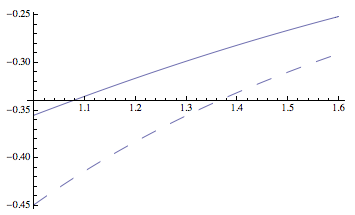}\\
Fig. 18.  $g_{V_3}(w)$ for the transition $0^- \to 1^+_{1/2}$ in the BT model at finite mass ($g_{V_3}(1) = -0.3534$) and in the heavy quark limit (dashed line). 

\vskip 0.5 truecm

\includegraphics[scale=1.]{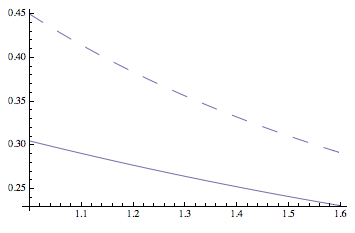}\\
Fig. 19.  $g_A(w)$ for the transition $0^- \to 1^+_{1/2}$ in the BT model at finite mass ($g_A(1) = 0.3030$) and in the heavy quark limit (dashed line). 

\vskip 0.5 truecm

\includegraphics[scale=1.]{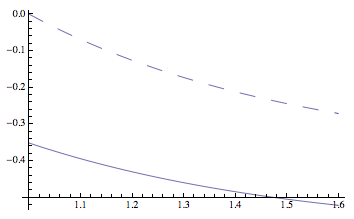}\\
Fig. 20.  $f_{V_1}(w)$ for the transition $0^- \to 1^+_{3/2}$ in the BT model at finite mass ($f_{V_1}(1) = -0.3567$) and in the heavy quark limit (dashed line).

\vskip 0.5 truecm

\includegraphics[scale=1.]{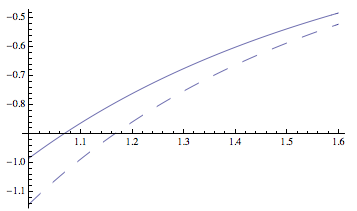}\\ 
Fig. 21.  $f_{V_2}(w)$ for the transition $0^- \to 1^+_{3/2}$ in the BT model at finite mass ($f_{V_2}(1) = -0.9720$) and in the heavy quark limit (dashed line).

\vskip 0.5 truecm

\includegraphics[scale=1.]{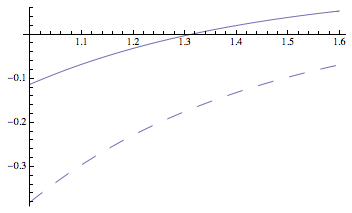}\\
Fig. 22.  $f_{V_3}(w)$ for the transition $0^- \to 1^+_{3/2}$ in the BT model at finite mass ($f_{V_3}(1) = -0.1090$) and in the heavy quark limit (dashed line). 

\vskip 0.5 truecm

\includegraphics[scale=1.]{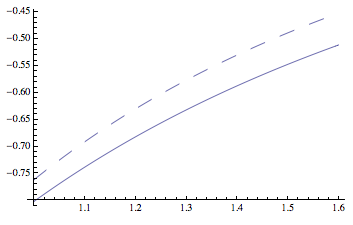}\\
Fig. 23.  $f_A(w)$ for the transition $0^- \to 1^+_{3/2}$ in the BT model at finite mass ($f_A(1) = -0.7964$) and in the heavy quark limit (dashed line).

\vskip 0.5 truecm

\includegraphics[scale=1.]{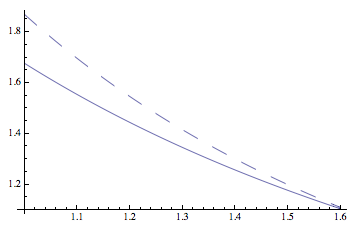}\\
Fig. 24.  $k_{A_1}(w)$ for the transition $0^- \to 2^+_{3/2}$ in the BT model at finite mass ($k_{A_1}(1) = 1.6756$) and in the heavy quark limit (dashed line).

\vskip 0.5 truecm

\includegraphics[scale=1.]{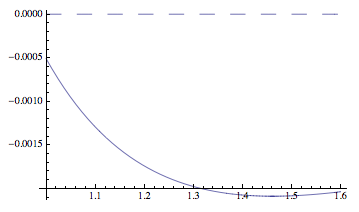}\\
Fig. 25.   $k_{A_2}(w)$ for the transition $0^- \to 2^+_{3/2}$ in the BT model at finite mass ($k_{A_2}(1) = -0.00311$) and in the heavy quark limit (dashed line).

\vskip 0.5 truecm

\includegraphics[scale=1.]{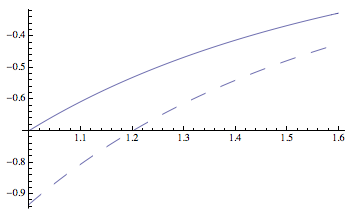}\\
Fig. 26.   $k_{A_3}(w)$ for the transition $0^- \to 2^+_{3/2}$ in the BT model at finite mass ($k_{A_3}(1) = -0.69823$) and in the heavy quark limit (dashed line).

\vskip 0.5 truecm

\includegraphics[scale=1.]{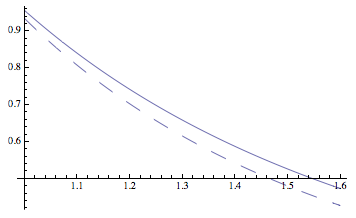}\\
Fig. 27.   $k_V(w)$ for the transition $0^- \to 2^+_{3/2}$ in the BT model at finite mass ($k_V(1) = 0.95574$) and in the heavy quark limit (dashed line).

\section {Branching ratios of $\overline{B} \to D^{(*)} \ell \nu, D^{**} \ell \nu, D^{(*)} \pi, D^{**} \pi$}

We now use formulas (\ref{E-1e})-(\ref{E-7e}) to compute the semileptonic branching ratios, and formula (\ref{E-8e}) to compute the pionic ones. 

\underline {At infinite mass}, only the form factors are computed in the heavy quark limit, while the kinematics contains the physical masses. One obtains, for the semileptonic modes :
$$BR(B \to D \ell \nu) = 2.022\ \% \qquad  \qquad \qquad \qquad \qquad \qquad \qquad$$ 
$$BR(B \to D^{*} \ell \nu) = 5.894\ \% \qquad  \qquad \qquad \qquad \qquad \qquad \qquad$$
$$BR(B \to D^{**}(0^+_{1/2}) \ell \nu) = 5.4 \times 10^{-4} \qquad  \qquad \qquad \qquad \qquad \ $$
$$BR(B \to D^{**}(1^+_{1/2}) \ell \nu) = 5.6 \times 10^{-4} \qquad  \qquad \qquad \qquad \qquad \ $$ 
$$BR(B \to D^{**}(1^+_{3/2}) \ell \nu) = 3.89 \times 10^{-3} \qquad  \qquad \qquad \qquad \ \ \ \ \ $$
\bea
\label{8.1e}
BR(B \to D^{**}(2^+_{3/2}) \ell \nu) = 6.04 \times 10^{-3} \qquad  \qquad \qquad \ \ \ \  \
\eea

\noindent and for the corresponding pionic decays :
$$BR(B \to D \pi) = 3.73 \times 10^{-3} \qquad  \qquad \qquad \qquad \qquad \qquad \ \ $$ 
$$BR(B \to D^{*} \pi) = 3.86 \times 10^{-3} \qquad  \qquad \qquad \qquad \qquad \qquad \ \ $$
$$BR(B \to D^{**}(0^+_{1/2}) \pi) = 1.5 \times 10^{-4} \qquad  \qquad \qquad \qquad \qquad \ $$ 
$$BR(B \to D^{**}(1^+_{1/2}) \pi) = 1.2 \times 10^{-4} \qquad  \qquad \qquad \qquad \qquad \ $$
$$BR(B \to D^{**}(1^+_{3/2}) \pi) = 1.25 \times 10^{-3} \qquad  \qquad \qquad \qquad \qquad $$
\bea
\label{8.2e}
BR(B \to D^{**}(2^+_{3/2}) \pi) = 1.19 \times 10^{-3} \qquad  \qquad \qquad \qquad \  
\eea

\noindent The pionic decays with form factors in the heavy quark limit have been compared to the Belle data \cite{BELLE D**PI} in ref. \cite{JLOR}.

On the other hand, \underline{at finite mass}  one has the following semileptonic BR :
$$BR(B \to D \ell \nu) = 2.354\ \% \qquad  \qquad \qquad \qquad \qquad \qquad \qquad$$ 
$$BR(B \to D^{*} \ell \nu) = 6.312\ \% \qquad  \qquad \qquad \qquad \qquad \qquad \qquad$$ 
$$BR(B \to D^{**}(0^+_{1/2}) \ell \nu) = 2.77 \times 10^{-3} \qquad  \qquad \qquad \qquad \ \ \ \ $$
$$BR(B \to D^{**}(1^+_{1/2}) \ell \nu) = 4.5 \times 10^{-4} \qquad  \qquad \qquad \qquad \qquad $$
$$BR(B \to D^{**}(1^+_{3/2}) \ell \nu) = 7.04 \times 10^{-3} \qquad  \qquad \qquad \qquad \ \ \ \ $$
\bea
\label{8.3e} 
BR(B \to D^{**}(2^+_{3/2}) \ell \nu) = 5.86 \times 10^{-3} \qquad  \qquad \qquad \ \ \ \ 
\eea

\noindent and the BR for pionic decays :
$$BR(B \to D \pi) = 0.469\ \% \qquad  \qquad \qquad \qquad \qquad \qquad \qquad$$ 
$$BR(B \to D^{*} \pi) = 0.476\ \% \qquad  \qquad \qquad \qquad \qquad \qquad \qquad$$ 
$$BR(B \to D^{**}(0^+_{1/2}) \pi) = 7.7 \times 10^{-4} \qquad  \qquad \qquad \qquad \qquad \ $$
$$BR(B \to D^{**}(1^+_{1/2}) \pi) = 1.1 \times 10^{-4} \qquad  \qquad \qquad \qquad \qquad \ $$
$$BR(B \to D^{**}(1^+_{3/2}) \pi) = 1.74 \times 10^{-3} \qquad  \qquad \qquad \qquad \ \ \ \ $$ 
\bea
\label{8.4e}
BR(B \to D^{**}(2^+_{3/2}) \pi) = 1.34 \times 10^{-3} \qquad  \qquad \qquad \ \ \ \ \            
\eea

Comparing the finite mass results with those in the heavy quark limit, we observe an enhancement in the case of the $0^+$ modes in both the semileptonic and pionic cases (about a factor 5), while the difference is moderate for the other decay modes. The enhancement for the $0^- \to 0^+$ transitions is due to a constructive interference in the decay rates between the two form factors $g_+(w)$ and $g_-(w)$. Of course, the magnitude of the enhancement is not trustable, since in this particular mode it is clearly related to the violation of the relation of Leibovich et al. In this case only two form factors contribute, and the subleading one should satisfy this relation.\par
In such a situation, it is not sensible to compare with the data of BaBar and Belle. A detailed discussion has been done recently of the experimental situation, compared with the BT model in the heavy quark limit and with the lattice results, in ref. \cite{LYP}.

\section{Discussion}

There cannot be a clear-cut conclusion for this work.\par

The Bakamjian-Thomas relativistic scheme was originally formulated to build states covariant under the Poincar\' e group. As shown in a number of papers, the BT relativistic quark model for hadron transitions is very satisfactory in the heavy quark limit. Indeed, in this limit current matrix elements are covariant, form factors exhibit Isgur-Wise scaling, and the Bjorken-Uraltsev sum rules are analytically satisfied.\par 
This model provides also a physical, phenomenological interpretation of a number of features of the heavy quark limit. One notorious example is the inequality $|\tau_{3/2}(1)| > |\tau_{1/2}(1)|$, that in the BT model is a spin effect due the Wigner rotation of the spin of the spectator light quark.\par
In the present paper we have tried to extend the BT model to finite mass, for the ground state transitions and for inelastic decays of the ground state to $L^P = 1^+$ excited states. However, at finite mass matrix elements are not covariant anymore and some unwanted results are not unexpected.\par
As exposed above, a convenient frame is the equal-velocity-frame, that we have adopted. On the theoretical side, to test the validity of the model at finite mass, at least the corrections at $O(1/m_Q)$ have to be compared with the rigorous results of HQET for these corrections.\par 
Among the latter, there are the consequences from HQET for the ground state case $0^- \to 0^-, 1^-$, i.e. Luke's theorem for the Lagrangian perturbations at zero recoil, and relations between the different Current perturbations for all $w$, established by Falk and Neubert. We have checked that these rigorous results of HQET are perfectly satisfied in the BT model at finite mass, even for all $w$ in the case of Current perturbations. In particular, the interesting relation between leading and subleading quantities $L_5(w) = - \overline{\Lambda}\xi(w)$ is analytically fulfilled.\par 
Other rigorous results of HQET at $O(1/m_Q)$ concern the values of the subleading form factors at zero recoil for transitions of the ground state to positive parity mesons $0^- \to 0^+, 1^+_{1/2}, 1^+_{3/2}$. These constraints on the subleading form factors, formulated by Leibovich, Ligeti, Stewart and Wise, exhibit a certain pattern in $1/m_Q\ (Q = b, c)$ and are proportional to the level spacings $\Delta E_j\ (j = 1/2, 3/2)$. In the model, the pattern in $1/m_Q\ (Q = b, c)$ is obtained in the model, but the proportionality to $\Delta E_j$ does not hold. This feature has an important numerical impact on the subleading form factor for the decays $\overline{B}(0^-) \to D^{**}(0^+) \ell \nu$, $\overline{B}(0^-) \to D^{**}(0^+) \pi$ resulting in a spurious enhancement of these decay rates.\par
As our analysis shows, in a formulation of relativistic quark models for such meson form factors, it is crucial to ensure that the relations of Leibovich et al. are satisfied  in the heavy quark expansion. It seems to us that to implement these relations is not obvious, and one should investigate whether they hold in other formulations of relativistic quark models.\par 
The BT scheme is not a particular model, but a very general framework. In fact, a framework quite similar to the one of BT is at the basis of the light front relativistic quark models \cite{KP,T,CGNPSS}. The same inelastic transitions $L = 0$ to $L = 1$ have been studied in the light front models of Cheng et al. \cite{CHENG}. But, to our knowledge, the problem of the identities of Leibovich et al. has not been evoked in this study.\par 
A similar approach, but based on the point form of BT, has been developped by M. G\'omez-Rocha and W. Schweiger \cite{GS}. These authors compute the form factors for the ground state transitions $\overline{B} \to D^{(*)} \ell \nu$. It would be very interesting to know if within their formalism they could confirm or not our results for the transitions $L = 0 \to L = 1$. \par
On the other hand, this problem has been clearly raised by Ebert et al. \cite{EFG}. In their relativistic quark model the identities are not automatically fulfilled, but imposed by a choice of the parameters of the potential. In our BT scheme, this latter possibility is clearly excluded.\par
For our part, one would wish to solve the problem of inelastic form factors in a general way through a fully covariant approach. This approach exists in the Bakamjian-Thomas framework in the heavy quark limit, but is lacking for the moment at finite mass.    

\vskip 0.5 truecm

{\large \bf Appendix A. Form factors in terms of matrix elements}

\vskip 0.5 truecm

From the definitions of Section 2, one can isolate the different form factors by introducing convenient four-vectors. The form factors for the $0^- \to 0^-$ transitions are simply given by
\bea
\label{A-1e}
\sqrt{m_Bm_D}\ h_+(w) = {<D(v')|(v+v').V|B(v)> \over 2(1+w)}
\eea
\bea
\label{A-2e}
\sqrt{m_Bm_D}\ h_-(w) = {<D(v')|(v-v').V|B(v)> \over 2(1-w)}
\eea

\noindent To isolate the $B \to D^{*}$ form factors we need to consider the longitudinal and transverse polarization four-vectors. Assuming the motion along the $Oz$ axis, we can adopt the following four-vectors :
\bea
\label{A-3e}
v = (v^0,0,0,v^z) \qquad \qquad \qquad \qquad v' = (v'^0,0,0,v'^z)
\eea
\bea
\label{A-4e}
\epsilon'^{(L)} = (v'^z,0,0,v'^0) \qquad \qquad \qquad \epsilon'^{(T)} = (0,1,0,0)
\eea

Then, the different form factors for the $0^- \to 1^-$ transitions are given by the expressions :
\bea
\label{A-5e}
\sqrt{m_Bm_{D^{*}}}\ h_V(w) = - {<D^{*(T)}(v')|i\epsilon_{\mu\nu\rho\sigma}\ \epsilon'^{(T)\nu} v'^\rho v^\sigma V^\mu |B(v)> \over w^2-1}
\eea
\bea
\label{A-6e}
\sqrt{m_Bm_{D^{*}}}\ h_{A_1}(w) = - {<D^{*(T)}(v')|\epsilon'^{(T)}.A|B(v)> \over w+1}
\eea
\bea
\label{A-7e}
\sqrt{m_Bm_{D^{*}}}\ h_{A_2}(w) = - {<D^{*(L)}(v')|\epsilon'^{(L)}.A|B(v)> - <D^{*(T)}(v')|\epsilon'^{(T)}.A|B(v)>  \over (\epsilon'^{(L)}.v)^2}
\eea
\bea
\label{A-8e}
&&\sqrt{m_Bm_{D^{*}}}\ h_{A_3}(w) = - {<D^{*(L)}(v')|v'.A|B(v)> \over (\epsilon'^{(L)}.v)} \nn \\
&& + {w (<D^{*(L)}(v')|\epsilon'^{(L)}.A|B(v)> - <D^{*(T)}(v')|\epsilon'^{(T)}.A|B(v)>) \over (\epsilon'^{(L)}.v)^2}
\eea

Similar relations for the form factors of the transitions to excited states can be obtained from the definitions (\ref{3.17e})-(\ref{3.23e}) :\par
\bea
\label{A-9e}
\sqrt{m_Bm_D}\ g_+(w) = {<D^{(1/2)}(0^+)(v')|(v+v').A|B(v)> \over 2(1+w)}
\eea
\bea
\label{A-10e}
\sqrt{m_Bm_D}\ g_-(w) = {<D^{(1/2)}(0^+)(v')|(v-v').A|B(v)> \over 2(1-w)}
\eea

\bea
\label{A-11e}
\sqrt{m_Bm_{D^{**}}}\ g_A(w) = - {<D^{(1/2)}(1^{+})^{(T)}(v')|i \epsilon_{\mu\nu\rho\sigma}\ \epsilon'^{(T)\nu} v'^\rho v^\sigma A^\mu |B(v)> \over w^2-1}
\eea
\bea
\label{A-12e}
\sqrt{m_Bm_{D^{**}}}\ g_{V_1}(w) = - <D^{(1/2)}(1^+)^{(T)}(v')|\epsilon'^{(T)}.V|B(v)>
\eea
$$\sqrt{m_Bm_{D^{**}}}\ g_{V_2}(w) = \qquad \qquad \qquad \qquad \qquad \qquad \qquad \qquad \qquad \qquad $$
\bea
\label{A-13e}
{<D^{(1/2)}(1^+)^{(L)}(v')|\epsilon'^{(L)}.V|B(v)> - <D^{(1/2)}(1^+)^{(T)}(v')|\epsilon'^{(T)}.V|B(v)> \over (\epsilon'^{(L)}.v)^2}
\eea
\bea
\label{A-14e}
\sqrt{m_Bm_{D^{**}}}\ g_{V_3}(w) = {<D^{(1/2)}(1^+)^{(L)}(v')|v'.V|B(v)> \over (\epsilon'^{(L)}.v)}
\eea
$$- {w (<D^{(1/2)}(1^+)^{(L)}(v')|\epsilon'^{(L)}.V|B(v)> - <D^{(1/2)}(1^+)^{(T)}(v')|\epsilon'^{(T)}.V|B(v)>) \over (\epsilon'^{(L)}.v)^2}$$ 

\noindent and similar formulas for the form factors $f_A(w)$, $f_{V_1}(w)$, $f_{V_2}(w)$ and $f_{V_3}(w)$ for the $D^{(3/2)}(1^+)$ state. Notice also that in the definition of the axial current matrix element for the ground state $D^*$, the form factor $h_{A_1}(w)$ is affected by a factor $(w+1)$, that does not appear in the corresponding definition of the vector form factors $g_{V_1}(w)$, $f_{V_1}(w)$ for the $1^+$ states.\par

To isolate the different form factors for the $2^+$ states, let us first write the corresponding tensor polarizations $\epsilon'^{(\lambda)}_{\mu\nu}$, that are symmetric $\epsilon'^{(\lambda)}_{\mu\nu} = \epsilon'^{(\lambda)}_{\nu\mu}$, traceless $g^{\mu \nu}\epsilon'^{(\lambda)}_{\mu \nu} = 0$ and transverse $v'^{\mu}\epsilon'^{(\lambda)}_{\mu\nu} = v'^{\nu}\epsilon'^{(\lambda)}_{\mu\nu} = 0$. The polarization tensors we are interested in (the currents are vectors) can be written as
$$\epsilon'^{(0)}_{\mu\nu} = {1 \over \sqrt{6}} \left[\epsilon'^{(+1)}_{\mu}\epsilon'^{(-1)}_{\nu}+2\epsilon'^{(0)}_{\mu}\epsilon'^{(0)}_{\nu}+\epsilon'^{(-1)}_{\mu}\epsilon'^{(+1)}_{\nu} \right ]$$ 
\bea
\label{A-20e}
\epsilon'^{(T)}_{\mu\nu} = {1 \over \sqrt{2}} \left[\epsilon'^{(T)}_{\mu}\epsilon'^{(0)}_{\nu}+\epsilon'^{(0)}_{\mu}\epsilon'^{(T)}_{\nu} \right ]
\eea

\noindent where $\epsilon'^{(T)}$ is the linear polarization vector (\ref{A-4e}), $\epsilon'^{(\lambda)}_\mu$ are the usual circular polarizations vectors ($\epsilon'^{(\lambda)}.\epsilon'^{(\lambda)} = -1, \epsilon'^{(\lambda)}.v' = 0$). In consistency with the motion along $Oz$ (\ref{A-3e}) we have
\bea
\label{A-21e}
\epsilon'^{(0)} = (v'^z,0,0,v'^0) \qquad \qquad \epsilon'^{(\pm1)} = \left(0,\mp{1 \over \sqrt{2}},- {i \over \sqrt{2}},0\right)
\eea

The different $2^+$ form factors will write, with the notation $\epsilon'^{(0)} = \epsilon'^{(L)}$,

\bea
\label{A-22e}
\sqrt{m_Bm_{D^{**}}}\ k_{A_1}(w) = -\sqrt{2}\ {<D^{(3/2)}(2^+)^{(T)}(v')|\epsilon'^{(T)}.A|B(v)> \over \epsilon'^{(L)}.v}
\eea
\bea
\label{A-23e}
\sqrt{m_Bm_{D^{**}}}\ k_{A_2}(w) \qquad \qquad \qquad \qquad \qquad \qquad \qquad \qquad \qquad \qquad 
\eea
$$= {\sqrt{3 \over 2} <D^{(3/2)}(2^+)^{(0)}(v')|\epsilon'^{(L)}.A|B(v)> - \sqrt{2} <D^{(3/2)}(2^+)^{(T)}(v')|\epsilon'^{(T)}.A|B(v)> \over (\epsilon'^{(L)}.v)^3}$$

\bea
\label{A-24e}
\sqrt{m_Bm_{D^{**}}}\ k_{A_3}(w) = \sqrt{3 \over 2}\ {<D^{(3/2)}(2^+)^{(0)}(v')|v'.A|B(v)> \over (\epsilon'^{(L)}.v)^2} \qquad \qquad \qquad
\eea
$$ - w {\sqrt{3 \over 2} <D^{(3/2)}(2^+)^{(0)}(v')|\epsilon'^{(L)}.A|B(v)> - \sqrt{2} <D^{(3/2)}(2^+)^{(T)}(v')|\epsilon'^{(T)}.A|B(v)> \over (\epsilon'^{(L)}.v)^3}$$

\bea
\label{A-25e}
\sqrt{m_Bm_{D^{**}}}\ k_V(w) = - \sqrt{2}\ {<D^{*(T)}(v')|i\epsilon_{\mu\nu\rho\sigma} V^\mu \epsilon'^{(T)\nu} v'^\rho v^\sigma|B(v)> \over (w^2-1)(\epsilon'^{(L)}.v)}
\eea

\vskip 0.5 truecm

{\large \bf Appendix B. Wave functions in the heavy quark limit in the GI model}

\vskip 0.5 truecm

We have computed the ground state wave function $j^P = {1 \over 2}^-$ by expanding it in a truncated harmonic oscillator basis

\bea
\label{B-1e}
\varphi_{{1 \over 2}^-}(\vec{k}) = \sum_{n=0}^{n=15} C^{{1 \over 2}^-}_n (-1)^n (4\pi)^{3/4}2^n\ \sqrt{{(n!)^2 \over (2n+1)!}} {1 \over \beta^{3/2}}\ L{_n^{1/2}}\left({\vec{k}^2 \over {\beta^2}}\right) \rm{exp}\left(-{\vec{k}^2 \over {2 \beta^2}}\right)
\eea

With the parameters
\bea
\label{B-2e}
m_1 = 10^4 \ GeV \qquad \qquad m_2 = 0.220 \ GeV \qquad \qquad \beta = 0.5\ GeV
\eea
\noindent one gets the coefficients
\bea
\label{B-3e}
&&C^{{1 \over 2}^-}_{0,...15} = (0.9793537, 0.1176603, 0.1468293, 4.3721687\times 10^{-2}, \nn \\  
&& 4.8045449\times 10^{-2}, 2.0475958\times 10^{-2}, 2.1334046\times 10^{-2}, \nn \\
&& 1.0961787\times 10^{-2}, 1.1114890\times 10^{-2}, 6.3780537\times 10^{-3}, \nn \\
&& 6.3600712\times 10^{-3}, 3.9184764\times 10^{-3}, 3.8404907\times 10^{-3}, \nn \\
&& 2.4935019\times 10^{-3}, 2.3138365\times 10^{-3}, 1.6319989\times 10^{-3})
\eea

Similarly, one gets the following wave function for the lowest ${1 \over 2}^+$ state :
\bea
\label{B-4e}
\varphi_{{1 \over 2}^+}(\vec{k}) = \sum_{n=0}^{n=15} C^{{1 \over 2}^+}_n (-1)^n (4\pi)^{3/4}2^{n+1}\ \sqrt{{n!(n+1)! \over (2n+3)!}} {|\vec{k}| \over \beta^{5/2}}\ L{_n^{3/2}}\left({\vec{k}^2 \over {\beta^2}}\right) \rm{exp}\left(-{\vec{k}^2 \over {2 \beta^2}}\right)
\eea

\noindent with the following coefficients
\bea
\label{B-5e}
&&C^{{1 \over 2}^+}_{0,...15} = (0.9797808, 0.1129152, 0.1477815, 4.7028150 \times 10^{-2}, \nn \\
&&4.4749252 \times 10^{-2},  2.2688832 \times 10^{-2}, 1.8693443 \times 10^{-2}, \nn \\
&&1.2282215 \times 10^{-2}, 9.3433624 \times 10^{-3}, 7.2159977 \times 10^{-3},  \nn \\
&&5.1802760 \times 10^{-3}, 4.5010597 \times 10^{-3}, 3.0235867 \times 10^{-3}, \nn \\
&&2.9367937 \times 10^{-3}, 1.7230053 \times 10^{-3}, 1.9955065 \times 10^{-3})
\eea

And the wave function for the lowest ${3 \over 2}^+$ state :
\bea
\label{B-6e}
\varphi_{{3 \over 2}^+}(\vec{k}) = \sum_{n=0}^{n=15} C^{{3 \over 2}^+}_n (-1)^n (4\pi)^{3/4}2^{n+1}\ \sqrt{{n!(n+1)! \over (2n+3)!}} {|\vec{k}| \over \beta^{5/2}}\ L{_n^{3/2}}\left({\vec{k}^2 \over {\beta^2}}\right) \rm{exp}\left(-{\vec{k}^2 \over {2 \beta^2}}\right)
\eea

\noindent with the coefficients
\bea
\label{B-7e}
&&C^{{3 \over 2}^+}_{0,...15} = (0.9878460, 1.0599474 \times 10^{-2}, 0.1471102, 9.8141907 \times 10^{-3}, \nn \\
&&4.3046847 \times 10^{-2},  5.8332058 \times 10^{-3}, 1.7356267 \times 10^{-2}, \nn \\
&&3.4403985 \times 10^{-3}, 8.4537473 \times 10^{-3}, 2.0915067 \times 10^{-3},  \nn \\
&&4.6376493 \times 10^{-3}, 1.3029705 \times 10^{-3}, 2.7383780 \times 10^{-3}, \nn \\
&&8.2387996 \times 10^{-4}, 1.6385724 \times 10^{-3}, 5.3599390 \times 10^{-4})
\eea

The set of wave functions (\ref{A-1e})(\ref{A-3e})(\ref{A-5e}) are all normalized according to 
\bea
\label{B-8e}
&&\int  {d \vec{k} \over (2 \pi)^3}\ |\varphi(\vec{k})|^2\ = 1
\eea 

\vskip 0.5 truecm

\noindent {\large \bf Appendix C. Wave functions in the GI model at finite mass}

\vskip 0.5 truecm

At finite mass, the wave functions are parametrized by
\bea
\label{C-1e}
\varphi_{J^-}(\vec{k}) = \sum_{n=0}^{n=15} C^{J^-}_n (-1)^n (4\pi)^{3/4}2^n\ \sqrt{{(n!)^2 \over (2n+1)!}} {1 \over \beta^{3/2}}\ L{_n^{1/2}}\left({\vec{k}^2 \over {\beta^2}}\right) \rm{exp}\left(-{\vec{k}^2 \over {2 \beta^2}}\right)
\eea

\noindent for the ground states $J = 0, 1$, and by 
\bea
\label{C-2e}
\varphi_{J_j^+}(\vec{k}) = \sum_{n=0}^{n=15} C^{J_j^+}_n (-1)^n (4\pi)^{3/4}2^{n+1}\ \sqrt{{n!(n+1)! \over (2n+3)!}} {|\vec{k}| \over \beta^{5/2}}\ L{_n^{3/2}}\left({\vec{k}^2 \over {\beta^2}}\right) \rm{exp}\left(-{\vec{k}^2 \over {2 \beta^2}}\right)
\eea

\noindent with $J_j = 0_{1/2}, 1_{1/2}, 1_{3/2}, 2_{3/2}$.\par

 The pseudoscalar $B$ meson wave function is common to all intial states that we are considering. In the GI model, the mass parameters that fit the data for $B$ mesons are
\bea
\label{C-3e}
m_1=4.977\ GeV \qquad \qquad m_2 = 0.220\ GeV \qquad \qquad \beta = 0.5\ GeV
\eea
\noindent and the wave function coefficients are :
\bea
\label{C-4e}
&&C^{B(0^-)}_{0,...15} = (0.9690171, 0.1531175, 0.1649211, 6.2490419\times 10^{-2}, \nn \\  
&& 6.0360532\times 10^{-2}, 3.1558599\times 10^{-2}, 2.9348362\times 10^{-2}, \nn \\
&& 1.7991375\times 10^{-2}, 1.6438706\times 10^{-2}, 1.1053351 \times 10^{-2}, \nn \\
&& 9.9637937\times 10^{-3}, 7.1392222\times 10^{-3}, 6.2874621\times 10^{-3}, \nn \\
&& 4.7953418\times 10^{-3}, 3.8834463\times 10^{-3}, 3.5072465\times 10^{-3})
\eea

For the different charmed $D$ mesons, the spectrum is described using the parameters
\bea
\label{C-5e}
m_1 = 1.628\ GeV \qquad \qquad m_2 = 0.220\ GeV \qquad \qquad \beta = 0.5\ GeV
\eea
\noindent and the coefficients of the expansions (\ref{B-1e}) and (\ref{B-2e}) for the various quantum numbers are given by
\bea
\label{C-6e}
&&C^{D(0^-)}_{0,...15} = (0.9600527, 0.1799335, 0.1767118, 7.6031193\times10^{-2}, \nn \\  
&& 6.8335488\times10^{-2}, 3.9312087\times10^{-2}, 3.4507290\times10^{-2}, \nn \\
&& 2.2833729\times10^{-2}, 1.9844856\times10^{-2}, 1.4270671\times10^{-2}, \nn \\
&& 1.2243154\times10^{-2}, 9.4011556\times10^{-3}, 7.7781440\times10^{-3}, \nn \\
&& 6.5341271\times10^{-3}, 4.6821525\times10^{-3}, 5.1816395\times10^{-3})
\eea
\bea
\label{C-7e}
&&C^{D(1^-)}_{0,...15} = (0.9894823, 4.9004469\times 10^{-2}, 0.1262952, 2.2102771\times 10^{-2}, \nn \\  
&& 3.5959065\times 10^{-2}, 1.0480723\times 10^{-2}, 1.4237838\times 10^{-2}, \nn \\
&& 5.380643\times 10^{-3}, 6.7386944\times 10^{-3}, 2.9314966\times 10^{-3}, \nn \\
&& 3.5624162\times 10^{-3}, 1.6525286\times 10^{-3}, 2.0363566\times 10^{-3}, \nn \\
&& 9.2892419\times 10^{-4}, 1.2249013\times 10^{-3}, 5.2336301\times 10^{-4})
\eea
\bea
\label{C-8e}
&&C^{D(0^+_{1/2})}_{0,...15} = (0.9848158, 5.2615825\times 10^{-2}, 0.1519192, 3.4893338\times 10^{-2}, \nn \\  
&& 4.5274679\times 10^{-2}, 1.9408170\times 10^{-2}, 1.8440058\times 10^{-2}, \nn \\
&& 1.1052819\times 10^{-2}, 8.9459708\times 10^{-3}, 6.5899095\times 10^{-3}, \nn \\
&& 4.7909911\times 10^{-3}, 4.1152863\times 10^{-3}, 2.6809596\times 10^{-3}, \nn \\
&& 2.7001044\times 10^{-3}, 1.4315639\times 10^{-3}, 1.9437365\times 10^{-3})
\eea
\bea
\label{C-9e}
&&C^{D(2^+_{3/2})}_{0,...15} = (0.9766909, -0.1460503, 0.1472010, -3.2608863\times 10^{-2}, \nn \\  
&& 3.9174896\times 10^{-2}, -9.5297368\times 10^{-3}, 1.4054954\times 10^{-2}, \nn \\
&& -3.5175697\times 10^{-3}, 6.1494103\times 10^{-3}, -1.5897267\times 10^{-3}, \nn \\
&& 3.0960441\times 10^{-3}, -8.5987244\times 10^{-4}, 1.7280004\times 10^{-3}, \nn \\
&& -5.3811091\times 10^{-4}, 1.0204369\times 10^{-3}, -3.2012642\times 10^{-4})
\eea

For the two states $D_1(1^+)$ and $D_2(1^+)$ the situation is more complicated because at finite mass they are not pure $j = {1 \over 2}$ or $j = {3 \over 2}$. From the GI model we find that each of these states has two components with $j = {1 \over 2}$ and $j = {3 \over 2}$.\par 
The two $j = {1 \over 2}$ and  $j = {3 \over 2}$ components of the $D_1(1^+)$ state, that is dominantly $j = {1 \over 2}$, are the following :
\bea
\label{C-10e}
&&C^{D_1(1^+)_{1/2}}_{0,...15} = (0.9750784, 4.2226720 \times 10^{-4}, 0.1388684, 1.9337032 \times 10^{-2}, \nn \\
&& 3.8017304 \times 10^{-2}, 1.3402707 \times 10^{-2}, 1.4626256 \times 10^{-2}, \nn \\
&& 8.3450762 \times 10^{-3}, 6.9152913 \times 10^{-3}, 5.2573497 \times 10^{-3}, \nn \\
&& 3.6921142 \times 10^{-3}, 3.424697 \times 10^{-3}, 2.0902278 \times 10^{-3}, \nn \\
&& 2.3228918 \times 10^{-3}, 1.1427986 \times 10^{-3}, 2.3432890 \times 10^{-3})
\eea
\bea
\label{C-11e}
&&C^{D_1(1^+)_{3/2}}_{0,...15} = (0.1630617, -1.7655547 \times 10^{-2}, 2.4015685 \times 10^{-2}, -4.2665031 \times 10^{-3}, \nn \\
&& 6.0282979 \times 10^{-3}, -1.6391013 \times 10^{-3}, 1.8471151 \times 10^{-3}, \nn \\
&& -9.1877274 \times 10^{-4}, 5.8052647 \times 10^{-4}, -6.3214857 \times 10^{-4}, \nn \\
&& 1.4512054 \times 10^{-4}, -4.7771402 \times 10^{-4}, -2.3710345 \times 10^{-6}, \nn \\
&& -3.8210297 \times 10^{-4}, -2.8517570 \times 10^{-5}, 1.2153198 \times 10^{-4})
\eea

\noindent Of course, the sum of the squared norms of the vectors (\ref{C-10e}) and (\ref{C-11e}) is normalized, $\sum_i [|C^{D_1(1^+)_{1/2}}_i|^2 + |C^{D_1(1^+)_{1/2}}_i|^2] = 1$.\par
On the other hand, the two $j = {1 \over 2}$ and  $j = {3 \over 2}$ components of the $D_2(1^+)$ state, that is dominantly $j = {3 \over 2}$, are the following :
\bea
\label{C-12e}
&&C^{D_2(1^+)_{1/2}}_{0,...15} = (-0.1633738, -4.8339165 \times 10^{-3}, -2.5603601 \times 10^{-2}, -5.8475351 \times 10^{-3}, \nn \\
&& -8.1983565 \times 10^{-3}, -3.9683113 \times 10^{-3}, -3.8034132 \times 10^{-3}, \nn \\
&& -2.6259074 \times 10^{-3}, -2.1547486 \times 10^{-3}, -1.7860712 \times 10^{-3}, \nn \\
&& -1.3476588 \times 10^{-3}, -1.2581809 \times 10^{-3}, -8.6974999 \times 10^{-4}, \nn \\
&& -9.28100700 \times 10^{-4}, -5.1979437 \times 10^{-4}, -7.6447322 \times 10^{-4})
\eea
\bea
\label{C-13e}
&&C^{D_2(1^+)_{3/2}}_{0,...15} = (0.9697097, 6.9651324 \times 10^{-2}, 0.1564430, -6.7026414 \times 10^{-3},\nn \\
&& 4.7112839 \times 10^{-2}, 2.2269301 \times 10^{-3}, 1.9536760 \times 10^{-2},\nn \\
&& 2.9314493 \times 10^{-3}, 9.8407984 \times 10^{-3}, 2.3789247 \times 10^{-3},\nn \\
&& 5.5826771 \times 10^{-3}, 1.7778778 \times 10^{-3}, 3.3748336 \times 10^{-3},\nn \\
&& 1.3486697 \times 10^{-3}, 1.9880311 \times 10^{-3}, 1.2300351 \times 10^{-3})
\eea

\noindent The sum of the squared norms of the vectors (\ref{C-12e}) and (\ref{C-13e}) is normalized as expected, $\sum_i [|C^{D_2(1^+)_{1/2}}_i|^2 + |C^{D_2(1^+)_{1/2}}_i|^2] = 1$.\par
 The wave functions of $D_1(1^+)$ and $D_2(1^+)$ must be orthogonal. The spin and orbital angular momentum parts of the wave functions $D_i(1^+)_{1/2}$ and $D_i(1^+)_{3/2}$ ($i = 1, 2$) are orthogonal. For the scalar product between $|D_1(1^+) >$ and $|D_2(1^+) >$ we are then left with the sum of products of the radial functions for given $j = {1 \over 2}$ and $j = {3 \over 2}$ that, from (\ref{C-10e})-(\ref{C-13e}), indeed vanishes :
\bea
\label{C-14e}
<D_1(1^+)|D_2(1^+)>\ \propto \sum_i [ C^{D_1(1^+)_{1/2}}_i C^{D_2(1^+)_{1/2}}_i + C^{D_1(1^+)_{3/2}}_i C^{D_2(1^+)_{3/2}}_i ] = 0
\eea

\vskip 0.5 truecm

{\large \bf Appendix D. A set of collinear frames}

\vskip 0.5 truecm

We have seen above that the current matrix elements in the BT model are covariant in the heavy quark limit. However, the subleading corrections in $1/m_Q$ are dependent on the frame. We  consider a family of collinear frames, with the mesons moving along the $Oz$ axis :
\bea
\label{D-1e}
v = (v^0, 0, 0, v^z) \qquad \qquad v' = (v'^0, 0, 0, v'^z)
\eea

\noindent going continously between the $B$ meson rest frame through the final $D$ meson rest frame. These frames can be labeled by a parameter $\alpha$, with $0 \leq \alpha \leq 1$ :
\bea
\label{D-2e}
(1-\alpha) v^z+\alpha v'^z = 0
\eea

\noindent The $B$ and the $D$ meson rest frames correspond respectively to $\alpha = 0$ and $\alpha = 1$, while the intermediate equal velocity frame (EVF), in which the spatial velocities are equal in modulus ($v^0 = v'^0, v^z = -v'^z$) corresponds to the value $\alpha = {1 \over 2}$.  

In terms of this parameter and of the variable $w = v.v'$, the four-vectors (\ref{C-1e}) then write
\bea
\label{D-3e}
&&v = \left( \sqrt{1 + {\alpha^2(w^2-1) \over \alpha^2+2\alpha(1-\alpha)w+(1-\alpha)^2}}, 0, 0, - \sqrt{{\alpha^2(w^2-1) \over \alpha^2+2\alpha(1-\alpha)w+(1-\alpha)^2}} \right) \nn \\
&&v' = \left( \sqrt{1 + {(1-\alpha)^2(w^2-1) \over \alpha^2+2\alpha(1-\alpha)w+(1-\alpha)^2}}, 0, 0, \sqrt{{(1-\alpha)^2(w^2-1) \over \alpha^2+2\alpha(1-\alpha)w+(1-\alpha)^2}} \right) \nn \\ 
\eea

\vskip 0.5 truecm

{\large \bf Appendix E. Formulas for the decay rates}

\vskip 0.5 truecm

The differential rates can be expressed in terms of the helicity amplitudes under the form
\bea
\label{E-1e}
{d\Gamma \over dw} = {G_F^2m_B^5 \over 48 \pi^3}\  {|V_{cb}|}^2\ r^3\ \sqrt{w^2-1} \left( |H_+(w)|^2 + |H_-(w)|^2 +  |H_0(w)|^2 \right) 
\eea

\noindent where $r = {m_D \over m_B}$ ($m_D$ being the mass of the corresponding charmed meson) and the helicity amplitudes squared write, in the different cases :

\vskip 0.5 truecm

$\bullet$ $\overline{B} \to D \ell \nu$
\bea
\label{E-2e}
H_\pm= 0
\eea
$$|H_0(w)|^2 = (w^2-1) \left[ (1+r)h_+(w)-(1-r)h_-(w) \right]^2$$

\vskip 0.5 truecm

$\bullet$ $\overline{B} \to D^* \ell \nu$
\bea
\label{E-3e}
|H_\pm(w)|^2 = (1+r^2-2rw) \left[(w+1) h_{A_1}(w) \mp \sqrt{w^2-1}\ h_V(w) \right]^2
\eea
$$|H_0(w)|^2 = (w+1)^2 \left\{(w-r) h_{A_1}(w) - (w-1)\left[ rh_{A_2}(w) + h_{A_3}(w)\right]  \right\}^2$$

\vskip 0.5 truecm

$\bullet$ $\overline{B} \to D^{**}(0^+_{1/2}) \ell \nu$
\bea
\label{E-4e}
H_\pm= 0
\eea
$$|H_0(w)|^2 = (w^2-1) \left[ (1+r)g_+(w)-(1-r)g_-(w) \right]^2$$

\vskip 0.5 truecm

$\bullet$ $\overline{B} \to D^{**}(1^+_{1/2}) \ell \nu$
\bea
\label{E-5e}
|H_\pm(w)|^2 = (1+r^2-2rw) \left[g_{V_1}(w) \mp \sqrt{w^2-1}\ g_A(w) \right]^2
\eea
$$|H_0(w)|^2 = \left\{(w-r) g_{V_1}(w) + (w^2-1)\left[ rg_{V_2}(w) + g_{V_3}(w)\right]  \right\}^2$$

\vskip 0.5 truecm

$\bullet$ $\overline{B} \to D^{**}(1^+_{3/2}) \ell \nu$
\bea
\label{E-6e}
|H_\pm(w)|^2 = (1+r^2-2rw) \left[f_{V_1}(w) \mp \sqrt{w^2-1}\ f_A(w) \right]^2
\eea
$$|H_0(w)|^2 = \left\{(w-r) f_{V_1}(w) + (w^2-1)\left[ rf_{V_2}(w) + f_{V_3}(w)\right]  \right\}^2$$

\vskip 0.5 truecm

$\bullet$ $\overline{B} \to D^{**}(2^+_{3/2}) \ell \nu$
\bea
\label{E-7e}
|H_\pm(w)|^2 = {1 \over 2}\ (1+r^2-2rw) (w^2-1) \left[k_{A_1}(w) \mp \sqrt{w^2-1}\ k_V(w) \right]^2
\eea
$$|H_0(w)|^2 = {2 \over 3}\ (w^2-1) \left\{(w-r) k_{A_1}(w) + (w^2-1) \left[ rk_{A_2}(w) + k_{A_3}(w)\right]  \right\}^2$$

\vskip 0.5 truecm

Of course, in the preceding formulas the masses of the charmed mesons, and hence the parameter $r$, vary according to the considered state $D, D^*, D^{**}(0^+_{1/2})$, $D^{**}(1^+_{1/2}), D^{**}(1^+_{3/2})$ or $D^{**}(2^+_{3/2})$. Remember also that the form factor $h_{A_1}(w)$ is affected by a factor $(w+1)$, that does not appear in the corresponding definition of the form factors $g_{V_1}(w)$, $f_{V_1}(w)$ for the $1^+$ states and also the form factors $h_{A_2}(w)$ and $h_{A_3}(w)$ are affected by a minus sign, contrarily to the definitions of  $g_{V_2}(w)$, $f_{V_2}(w)$ and  $g_{V_3}(w)$, $f_{V_3}(w)$ for the $1^+$ states, as we see in the definitions (\ref{3.16e})-(\ref{3.23e}).\par

The decays rates for pionic decays read :
\bea
\label{E-8e}
\Gamma_\pi = {3 \pi^2 |V_{ub}|^2 a_1^2 f_\pi^2 \over m_B m_D} \left({d\Gamma_{sl} \over dw}\right)_{w_{max}} \qquad \qquad \left(w_{max} = {{m_B^2+m_D^2} \over {2 m_B m_D}}\right) 
\eea

\noindent where $a_1 \simeq 1$ is a combination of Wilson coefficients, and $m_D$ is the mass of the corresponding charmed meson.

\vskip 1 truecm

\section*{Acknowledgements} \hspace*{\parindent} 
We acknowledge Damir Be\' cirevi\' c for discussions and useful advice, Roy Aleksan and Patrick Roudeau for remarks on the "$1/2$ vs. $3/2$ puzzle", and Vincent Mor\' enas for making available to us his very well written PhD Thesis. 


\begin{thebibliography}{99}

\bibitem{BT} B. Bakamjian and L.H. Thomas, Phys. Rev. {\bf 92}, 1300 (1953).
\bibitem{KP} B. D. Keister and W. N. Polyzou, Adv. Nucl. Phys. {\bf 20}, 225 (1991).
\bibitem{T} M. Terent'ev, Sov. J. Nucl. Phys. {\bf 24}, 106 (1976).
\bibitem{CGNPSS} F. Cardarelli, I. L. Grach, I. M. Nadoretskii, E. Pace, G. Sale and S. Simula, Phys. Lett. B {\bf 332}, 1 (1994).

\bibitem{LOPR-1} A. Le Yaouanc, L. Oliver, O. P\`ene and J.-C. Raynal, Phys. Lett. B {\bf 365}, 319 (1996).
\bibitem{MLOPR-1} V. Mor\'enas, A. Le Yaouanc, L. Oliver, O. P\`ene and J.-C. Raynal, Phys. Lett. B {\bf 386}, 315 (1996).

\bibitem{IW-1} N. Isgur, M. B. Wise, Phys. Lett. B {\bf 232}, 113 (1989); Phys. Lett. B {\bf 237}, 527 (1990).
\bibitem{IW-2} N. Isgur, M. B. Wise, Phys. Rev. D {\bf 43}, 819 (1991).

\bibitem{BJ} J. D. Bjorken, invited talk at Les Rencontres de la Vall ́ee d’Aoste, La Thuile, SLAC-PUB-5278, 1990.
\bibitem{UR} N. Uraltsev, Phys. Lett. B {\bf 501}, 86 (2001); N. Uraltsev, J. Phys. G {\bf 27}, 1081 (2001).

\bibitem{LOPR-2} A. Le Yaouanc, L. Oliver, O. P\`ene and J.-C. Raynal, Phys. Lett. B {\bf 386}, 304 (1996).
\bibitem{MLOPR-2} V. Mor\'enas, A. Le Yaouanc, L. Oliver, O. P\`ene and J.-C. Raynal, Phys. Lett. B {\bf 408}, 357 (1997).
\bibitem{LOPRM} A. Le Yaouanc, L. Oliver, O. P\`ene and J.-C. Raynal and V. Mor\'enas, Phys. Lett. B {\bf 520}, 25 (2001).

\bibitem{LOR-1} A. Le Yaouanc, L. Oliver and J.-C. Raynal, Phys. Rev. D {\bf 67}, 114009 (2003).
\bibitem{LOR-2} A. Le Yaouanc, L. Oliver and J.-C. Raynal, Phys. Lett. B {\bf 557}, 207 (2003). 
\bibitem{LOR-3} A. Le Yaouanc, L. Oliver and J.-C. Raynal, Phys. Rev. D {\bf 69}, 094022 (2004).

\bibitem{MLOPR-3} V. Mor\'enas, A. Le Yaouanc, L. Oliver, O. P\`ene and J.-C. Raynal, Phys. Rev. D {\bf 56}, 5668 (1997).
\bibitem{GI} S. Godfrey and N. Isgur, Phys. Rev. D {\bf 32}, 189 (1985).

\bibitem{MLOPR-4} V. Mor\'enas, A. Le Yaouanc, L. Oliver, O. P\`ene and J.-C. Raynal, Phys. Rev. D {\bf 58}, 114019 (1998).
\bibitem{LOPR-3} A. Le Yaouanc, L. Oliver, O. P\`ene and J.-C. Raynal, Phys. Lett. B {\bf 387}, 582 (1996). 
\bibitem{M} V. Mor\'enas, Th\` ese, Universit\' e Blaise Pascal, Clermont-Ferrand (1997).

\bibitem{BCLOR} D. Be\'cirevi\'c, E. Chang, A. Le Yaouanc, L. Oliver and J.-C. Raynal, Phys. Rev. D {\bf 84}, 054507 (2011). 

\bibitem{BIGI} I.I. Bigi, B. Blossier, A. Le Yaouanc, L. Oliver, O. P\`ene, J.-C. Raynal, A. Oyanguren, P. Roudeau, Eur. Phys. J. C {\bf 52} (2007). 

\bibitem{BABAR D** SL} B. Aubert et al., Phys. Rev. Lett. {\bf 101}, 261802 (2008).
\bibitem{BELLE D** SL} D. Liventsev et al., Phys. Rev. D {\bf 77}, 091503 (2008).

\bibitem{BLOSSIER} B. Blossier, M. Wagner and O. P\`ene, JHEP {\bf 0906}, 022 (2009).

\bibitem{BELLE D**PI} The BELLE Collab., K. Abe et al., hep-ex/0412072.

\bibitem{JLOR} F. Jugeau, A. Le Yaouanc, L. Oliver, J.-C. Raynal, Phys. Rev. D {\bf 72} (2005) 094010.

\bibitem{LYP} A. Le Yaouanc and O. P\`ene, in {\it In Memoriam} Nikolai Uraltsev, arXiv 1408.5104 [hep-ph], to be published in World Scientific Review (2014). 

\bibitem{FN} A. F. Falk and M. Neubert, Phys. Rev. D {\bf 47}, 2965 (1993).

\bibitem{LLSW} A. Leibovich, Z. Ligeti, I. Stewart and M. Wise, Phys. Rev. D {\bf 57}, 308 (1998).

\bibitem{L} M.E. Luke, Phys. Lett. B {\bf 252}, 447 (1990).

\bibitem{BSUV} I. Bigi, M. Shifman, N. Uraltsev and A. Vainshtein, Phys. Rev. D {\bf 52}, 196 (1995); Ann. Rev. Nucl. Part. Sci. {\bf 47}, 591 (1997).

\bibitem{CHENG} H.-Y. Cheng, C.-K. Chua and C.-W. Hwang, Phys. Rev. D {\bf 69}, 074025 (2004).

\bibitem{GS} M. G\'omez-Rocha and W. Schweiger, Phys. Rev. D {\bf 86}, 053010 (2012).

\bibitem{EFG} D. Ebert, R. Faustov and V. Galkin, Phys. Rev. D {\bf 61}, 014016 (2000); Phys. Rev. D {\bf 62}, 014032 (2000).

\end{thebibliography}
\end{document}